\documentclass[11pt,letterpaper,english,11pt,aps,letter,superscriptaddress,floatfix,pre]{revtex4}

\usepackage[T1]{fontenc}
\usepackage[latin1]{inputenc}
\usepackage{xcolor}
\usepackage{pdfcolmk}
\usepackage{babel}

\usepackage{amsmath}
\usepackage{graphicx}
\usepackage{amssymb}
\PassOptionsToPackage{normalem}{ulem}
\usepackage{ulem}
\usepackage[unicode=true, pdfusetitle,
 bookmarks=true,bookmarksnumbered=false,bookmarksopen=false,
 breaklinks=false,pdfborder={0 0 0},backref=false,colorlinks=false]
 {hyperref}

\makeatletter

\pdfpageheight\paperheight
\pdfpagewidth\paperwidth

\providecolor{lyxadded}{rgb}{1,0,0}
\providecolor{lyxdeleted}{rgb}{0,0,1}

\@ifundefined{textcolor}{}
{%
 \definecolor{BLACK}{gray}{0}
 \definecolor{WHITE}{gray}{1}
 \definecolor{RED}{rgb}{1,0,0}
 \definecolor{GREEN}{rgb}{0,1,0}
 \definecolor{BLUE}{rgb}{0,0,1}
 \definecolor{CYAN}{cmyk}{1,0,0,0}
 \definecolor{MAGENTA}{cmyk}{0,1,0,0}
 \definecolor{YELLOW}{cmyk}{0,0,1,0}
 }

\usepackage{ae,aecompl}

\renewcommand{\citet}[1]{\cite{#1}}

\makeatother

\begin{document}
\global\long\def\V#1{\boldsymbol{#1}}
\global\long\def\M#1{\boldsymbol{#1}}
\global\long\def\Set#1{\mathbb{#1}}

\global\long\def\D#1{\Delta#1}
\global\long\def\d#1{\delta#1}

\global\long\def\norm#1{\left\Vert #1\right\Vert }
\global\long\def\abs#1{\left|#1\right|}

\global\long\def\grad{\M{\nabla}}
\global\long\def\avv#1{\langle#1\rangle}
\global\long\def\av#1{\left\langle #1\right\rangle }

\global\long\def\ki{k}
\global\long\def\wi{\omega}

\title{\emph{Enhancement of Diffusive Transport by Nonequilibrium Thermal
Fluctuations}}

\author{Aleksandar Donev}

\email{donev@courant.nyu.edu}

\affiliation{Courant Institute of Mathematical Sciences, New York University,
New York, NY 10012}

\author{John B. Bell}

\affiliation{Center for Computational Science and Engineering, Lawrence Berkeley
National Laboratory, Berkeley, CA, 94720}

\author{Anton de la Fuente}

\affiliation{Department of Physics and Astronomy, San Jose State University, San
Jose, California, 95192}

\author{Alejandro L. Garcia}

\affiliation{Department of Physics and Astronomy, San Jose State University, San
Jose, California, 95192}
\begin{abstract}
We study the contribution of advection by thermal velocity fluctuations
to the effective diffusion coefficient in a mixture of two indistinguishable
fluids. The steady-state diffusive flux in a finite system subject
to a concentration gradient is enhanced because of long-range correlations
between concentration fluctuations and fluctuations of the velocity
parallel to the concentration gradient. The enhancement of the diffusive
transport depends on the system size $L$ and grows as $\ln(L/L_{0})$
in quasi-two dimensional systems, while in three dimensions it grows
as $L_{0}^{-1}-L^{-1}$, where $L_{0}$ is a reference length. The
predictions of a simple fluctuating hydrodynamics theory, closely
related to second-order mode-mode coupling analysis, are compared
to results from particle simulations and a finite-volume solver and
excellent agreement is observed. We elucidate the direct connection
to the long-time tail of the velocity autocorrelation function in
finite systems, as well as finite-size corrections employed in molecular
dynamics calculations. Our results conclusively demonstrate that the
nonlinear advective terms need to be retained in the equations of
fluctuating hydrodynamics when modeling transport in small-scale finite
systems.
\end{abstract}
\maketitle

\section{\label{Section_GiantFluct}Introduction}

Thermal fluctuations in non-equilibrium systems in which a constant
(temperature, concentration, velocity) gradient is imposed externally
exhibit remarkable behavior compared to equilibrium systems. Most
notably, external gradients can lead to \emph{enhancement} of thermal
fluctuations and to \emph{long-range} correlations between fluctuations
\citet{DSMC_Fluctuations_Shear,Mareschal:92,LongRangeCorrelations_MD,Zarate:04,FluctHydroNonEq_Book}.
This phenomenon can be illustrated by considering concentration fluctuations
in an isothermal mixture of two miscible fluids, subjected to a macroscopic
concentration gradient $\grad c$. The solution of the linearized
equations of fluctuating hydrodynamics shows that concentration and
density fluctuations exhibit long-range correlations, leading to a
power-law divergence of the static structure factors for small wavenumbers
$k$. When the species have different molecular masses and gravity
$\V g$ is present, the analysis predicts that fluctuations at wavenumbers
below $k_{g}\sim\left(\V g\cdot\grad c\right)^{1/4}$ are suppressed
\citet{GiantFluctuations_Nature,GiantFluctuations_Theory,GiantFluctuations_Universal,GiantFluctuations_Microgravity,GiantFluctuations_Cannell,FractalDiffusion_Microgravity},
where $c$ is the mass concentration of the heavier species. Similar
conclusions hold for fluctuations in a single-component fluid subject
to a stabilizing temperature gradient \citet{TemperatureGradient_Cannell}.

\begin{figure}[h]
\begin{centering}
\includegraphics[width=0.49\textwidth]{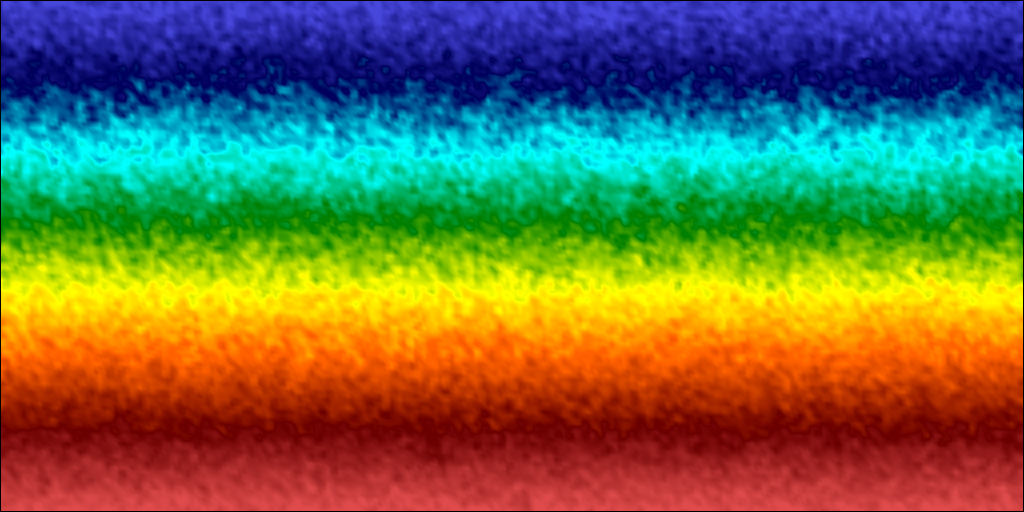}\hspace{0.25cm}\includegraphics[width=0.49\textwidth]{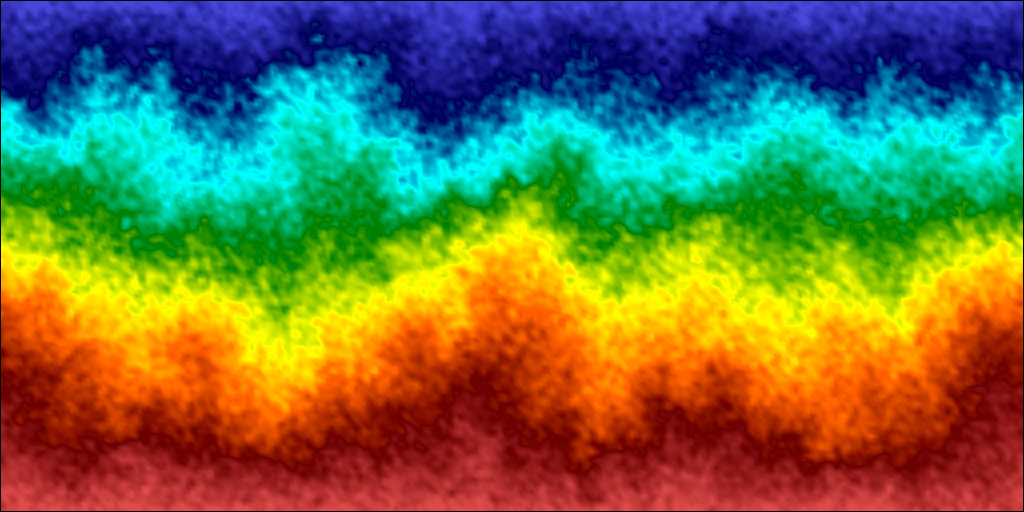}
\par\end{centering}

\begin{centering}
\vspace{0.25cm}\includegraphics[width=0.49\textwidth]{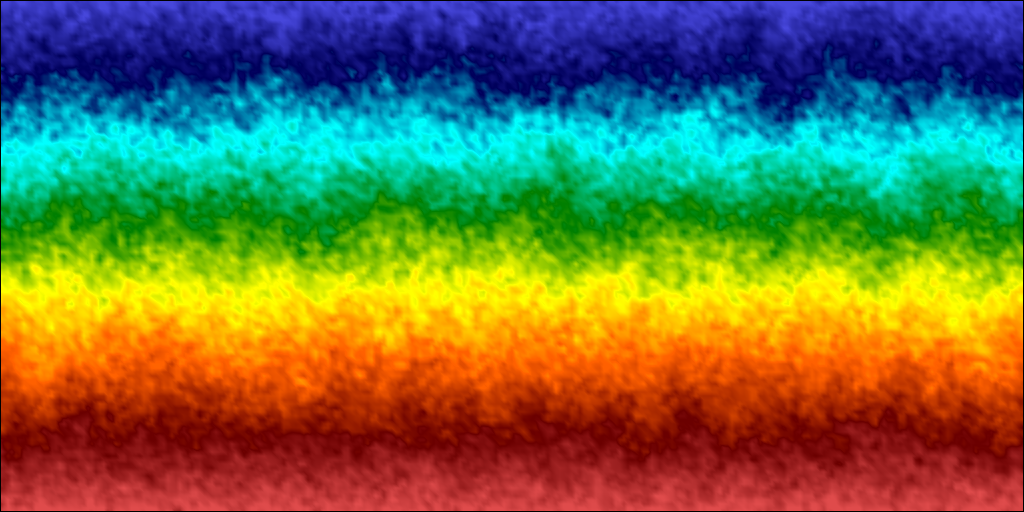}\hspace{0.25cm}\includegraphics[width=0.49\textwidth]{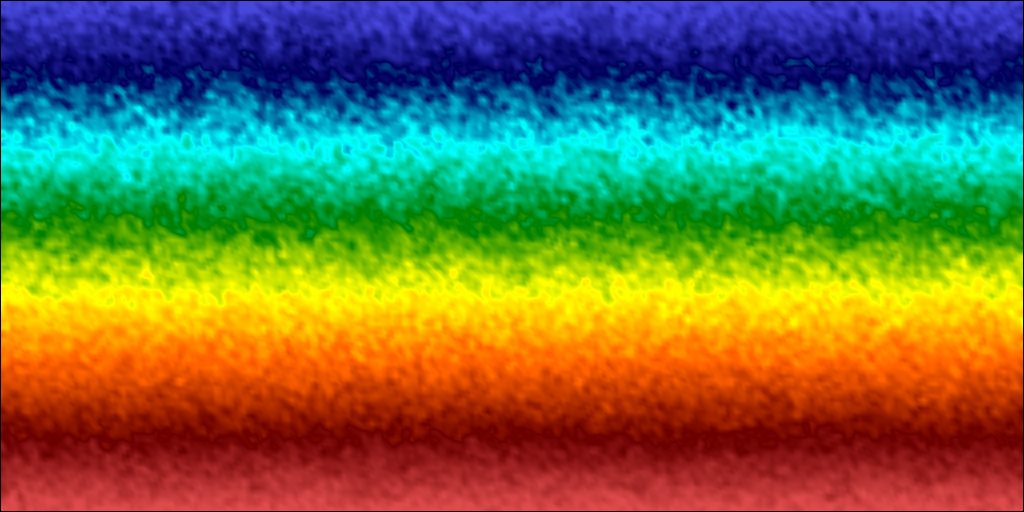}
\par\end{centering}

\caption{\label{fig:GiantFluctNeq}Snapshot of the local concentration for
a mixture of two ideal gases in a quasi-two-dimensional system, as
obtained from a long run of a compressible finite-volume solver which
includes thermal fluctuations \citet{LLNS_S_k}, with constant-temperature
walls placed at the top and bottom boundary. The particles of species
1 (red end of the color scale) are four times heavier than particles
of species 2 (blue end of the color scale), and all systems are in
mechanical equilibrium (i.e., the gravity force is balanced by the
pressure gradient). (\emph{Top left panel}) A system at \emph{equilibrium},
in the presence of gravity $g_{0}$, with the heavier particles sedimenting
toward the bottom according to the equilibrium Gibbs-Boltzmann distribution.
(\emph{Top right panel}) No gravity but a similar \emph{non-equilibrium}
concentration profile as in the top left panel is imposed via Dirichlet
boundary conditions at the top and bottom walls, showing giant concentration
fluctuations. (\emph{Bottom left panel}) The same boundary conditions
for the concentration are imposed as in the top right panel, but with
gravity $g=0.1\, g_{0}$, showing a suppression of the large-scale
giant fluctuations. (\emph{Bottom right panel}) The same boundary
conditions for the concentration are imposed as in the top right panel,
but with the same gravity $g=g_{0}$ as in the top left panel, showing
no enhancement of the fluctuations.}

\end{figure}

It is important to emphasize that this enhancement of large-scale
(small wavenumber) concentration fluctuations occurs because of the
non-equilibrium setting, and \emph{not} because of the concentration
gradient itself. Specifically, the enhancement is related to the \emph{dissipative}
flux through the system \citet{LLNS_Schmitz}, which is zero at thermodynamic
equilibrium. The top left panel of Fig. \ref{fig:GiantFluctNeq} shows
a snapshot of the concentration field for a two-dimensional system
in thermodynamic \emph{equilibrium} in which there is a concentration
gradient because of the sedimentation of the heavier species due to
gravity, but no enhancement of the fluctuations. By comparison, the
top right panel of the figure shows a \emph{non-equilibrium} system
with similar parameters but with an externally-imposed concentration
gradient (via the top and bottom wall boundary conditions) and no
gravity, revealing much-enhanced fluctuations (noise) and large-scale
features (clumping). If gravity is included in addition to the external
gradient, the total diffusive flux is reduced and large-scale fluctuations
(wavenumber $k\lesssim k_{g}$) are suppressed, as shown in the bottom
left panel of Fig. \ref{fig:GiantFluctNeq}. In fact, if the same
gravity as in the equilibrium case is imposed in addition to the external
gradient, the total diffusive flux is essentially zero and the system
is close to equilibrium again, giving no visible enhancement of the
concentration fluctuations over the equilibrium case, as shown in
the bottom right panel of Fig. \ref{fig:GiantFluctNeq}. These illustrative
numerical results were obtained using a finite-volume solver for compressible
fluctuating hydrodynamics \citet{LLNS_S_k}.

\begin{figure}[h]
\begin{centering}
\includegraphics[width=0.99\textwidth]{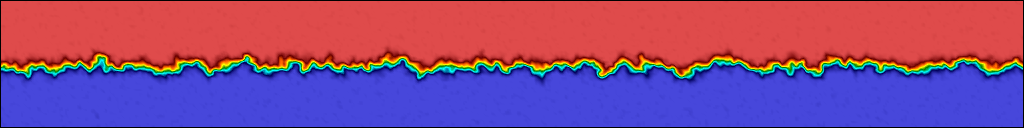}
\par\end{centering}

\begin{centering}
\includegraphics[width=0.99\textwidth]{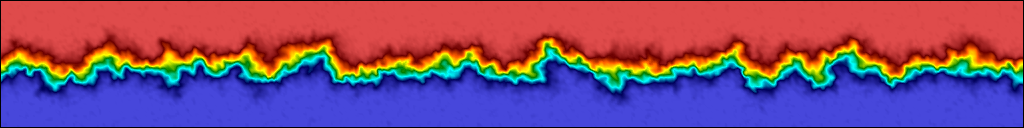}
\par\end{centering}

\begin{centering}
\includegraphics[width=0.99\textwidth]{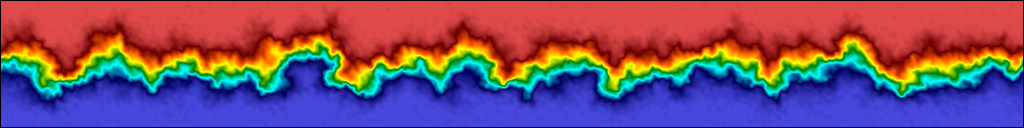}
\par\end{centering}

\caption{\label{fig:GiantFluctMixing}The \emph{rough} diffusive interface
between two miscible fluids at three points in time (top to bottom),
starting from an initially perfectly flat interface (phase separated
system), without gravity. Compare to the top right panel in Fig. \ref{fig:GiantFluctNeq}.}

\end{figure}

The enhancement of concentration fluctuations is even more dramatic
if the concentration gradient is at an interface, as in the study
of the early stages of diffusive mixing between initially separated
fluid components. As illustrated in Fig. \ref{fig:GiantFluctMixing},
the interface between the fluids, instead of remaining flat, develops
large-scale roughness that reaches a pronounced maximum until gravity
or boundary effects intervene. These \emph{giant fluctuations }\citet{GiantFluctuations_Theory,TemperatureGradient_Cannell,GiantFluctuations_ThinFilms}
during free diffusive mixing have been observed using light scattering
and shadowgraphy techniques \citet{GiantFluctuations_Nature,GiantFluctuations_Universal,GiantFluctuations_Microgravity,GiantFluctuations_Cannell,FractalDiffusion_Microgravity},
finding good but imperfect agreement between the predictions of a
simplified fluctuating hydrodynamic theory and experiments. In the
absence of gravity, the density mismatch between the two fluids does
not change the qualitative nature of the non-equilibrium fluctuations,
and in this work we focus on the case of two dynamically-indistinguishable
fluids.

The giant fluctuation phenomenon arises because of the appearance
of long-range correlations between concentration and velocity fluctuations
in the presence of a concentration gradient. Based on nonlinear fluctuating
hydrodynamic theory, it has been predicted that these correlations
give rise to fluctuation-renormalized transport coefficients at larger
scales \citet{DiffusionRenormalization_I,DiffusionRenormalization_II,ExtraDiffusion_Vailati}.
However, the predicted contribution from fluctuations to transport
at mesoscopic and macroscopic scales has only recently been computationally
observed and reported by the authors \citet{DiffusionRenormalization_PRL}.
This paper presents a detailed exposition of both the theoretical
prediction for the enhancement of diffusion and the numerical simulations
verifying these predictions. In particular, it is important to understand
how the effective transport coefficients depend on the length scale
of observation. This length scale may be related to the length $\D x$
of the finite-volume cells used in a fluctuating hydrodynamic solver,
or it may be related to the physical dimensions of a \emph{finite
system} such as a nano-channel transporting liquid or a nano-wire
transporting heat.

We consider diffusion in\emph{ }a mixture of dynamically identical
but labeled (as components 1 and 2) fluids \citet{SelfDiffusion_Linearity}
enclosed in a box of lengths $L_{x}\times L_{y}\times L_{z}$, in
the absence of gravity. Periodic boundary conditions are applied in
the $x$ (horizontal) and $z$ (depth) directions, and impermeable
constant-temperature walls are placed at the top and bottom boundaries.
A concentration gradient $\nabla\bar{c}=(c_{T}-c_{B})/L_{y}$ is imposed
along the $y$ axes by enforcing a constant mass concentration $c_{T}$
at the top wall and $c_{B}$ at the bottom wall. Because the fluids
are indistinguishable, concentration does not affect the fluid properties,
and the dynamics of the density, temperature and velocity fluctuations
remains as in thermodynamic equilibrium. In this sense, concentration
is \emph{passively transported} by thermal fluctuations, analogous
to diffusion of a passive tracer in a turbulent velocity field \citet{HomogeneousTurbulence_Batchelor,TurbulenceClosures_Majda}.
Note that for large $L_{y}$ the mass flux will be proportional to
the self-diffusion coefficient of a tagged particle, independent of
the magnitude of the gradient \citet{SelfDiffusion_Linearity}.

Since species are not changed in particle collisions, the diffusive
transport of particle label (concentration) can only occur via advective
motion of the particles between collisions. Kinetic theory shows that
at steady state the particles of a given species (denoted either with
a subscript or with a parenthesis superscript) have a non-zero macroscopic
momentum density $\bar{\V j}_{1}=\bar{\rho}_{1}\bar{\V v}_{1}=-\bar{\V j}_{2}=-\bar{\rho}_{2}\bar{\V v}_{2}$,
where $\rho$ denotes density and $\V v$ velocity. If the labeling
of the species is ignored, the system is at equilibrium and the overall
center-of-mass velocity vanishes, $\bar{\V v}=\bar{c}\bar{\V v}_{1}+\left(1-\bar{c}\right)\bar{\V v}_{2}=\V 0$.
More detailed kinetic theory \citet{BinaryMixKineticTheory,MultiFluidKineticTheory}
shows that the inter-species velocity $\bar{\V v}_{12}=\bar{\V v}_{1}-\bar{\V v}_{2}$
quickly relaxes to its equilibrium value,\begin{equation}
\bar{\V v}_{12}=-\frac{\chi\left(\grad\bar{c}\right)}{\bar{c}\left(1-\bar{c}\right)},\label{eq:v12_bar}\end{equation}
giving the Fickian diffusive flux for the mass concentration $c=\rho_{1}/\rho$
\citet{BinaryMixKineticTheory}, \[
\bar{\V j}_{1}=\bar{\rho}_{1}\bar{\V v}_{1}=\rho_{1}\bar{\V v}-\bar{\rho}\chi\left(\grad\bar{c}\right)=-\bar{\rho}\chi\left(\grad\bar{c}\right),\]
where $\chi$ is the mass diffusion coefficient. The local fluctuations
around the macroscopic mean, $\rho_{1}=\bar{\rho}_{1}+\d{\rho}_{1}$
and $\V v_{1}=\bar{\V v}_{1}+\d{\V v}_{1}$, can also make a non-trivial
contribution to the average mass flux,\begin{equation}
\av{\V j_{1}}=\av{\rho_{1}\V v_{1}}=-\bar{\rho}\chi\left(\grad\bar{c}\right)+\av{\left(\d{\rho}_{1}\right)\left(\d{\V v}_{1}\right)},\label{eq:av_j1}\end{equation}
if they are correlated, which in fact they are in the presence of
a concentration gradient.

The fluctuating hydrodynamics formalism \citet{Landau:Fluid,FluctHydroNonEq_Book}
is the most direct way to calculate steady-state correlations between
hydrodynamic variables, especially if a spatial Fourier transform
is used to separate different wavenumbers. Converting the correlations
from Fourier to real space requires integrating over all wavenumbers,
which gives qualitatively different results in two and in three dimensions,
as detailed in Section \ref{sec:Fluctuating-Hydrodynamics}. In two
dimensions the average mass flux, or effective diffusion coefficient,
is found to grow logarithmically with system size, while in three
dimensions an asymptotic {}``macroscopic'' value is reached for
sufficiently large systems.

As seen from (\ref{eq:av_j1}), the correlation that is needed is
that between the density and velocity of a given species, $\d{\rho}_{1}$
and $\d{\V v}_{1}$. However, in the standard {}``single-fluid''
hydrodynamic description of mixtures, unlike the little-understood
{}``two-fluid'' models \citet{BinaryMixKineticTheory,MultiFluidKineticTheory},
the individual species velocity $\V v_{1}$ (or equivalently, $\V v_{12}$)
is not maintained as an independent variable and instead only the
center-of-mass velocity $\V v$ appears \citet{FluctHydroNonEq_Book}.
In Section \ref{sec:Particle-Simulations} we report results from
simulations based on the Direct Simulation Monte Carlo (DSMC) particle
method \citet{DSMC_Bird,DSMCReview_Garcia}, in which we calculate
the spectral correlations between $\d{\rho}_{1}$ and $\d{\V v}_{1}$
and also the mass flux. The results presented in Section \ref{sub:Static-Factor}
indicate that the correlation between $\d{\rho}_{1}$ and $\d{\V v}_{1}$
is well-approximated by the prediction of the incompressible single-fluid
theory presented in Section \ref{sec:Fluctuating-Hydrodynamics}.
The effective diffusion coefficient is found to increase with system
size in accordance with the theory as well, as detailed in Section
\ref{sub:DSMC_Results}. For systems with aspect ratio close to unity
the use of the periodic (Fourier-based) theory is not appropriate
and the proper boundary conditions ought to be taken into account,
as we do by using a recently-developed compressible finite-volume
solver \citet{LLNS_S_k} in Section \ref{sub:FiniteHeight}. Very
good agreement is observed between the finite-volume simulations and
the particle results over a broad range of system sizes, once the
local diffusion coefficient that appears in the equations of fluctuating
hydrodynamics is adjusted to match particle data for a chosen reference
system. This locally-renormalized diffusion coefficient is also measured
in the particle simulations and found to match the theoretical predictions
reasonably well, as discussed in Sec. \ref{sub:chi_0}.

The present paper builds on an extensive prior literature on the renormalization
of the diffusion coefficient by hydrodynamic fluctuations and interactions.
In Section \ref{sec:VACF} we discuss connections to prior work in
more detail, and find that several different theoretical approaches
produce the same results as the very simple, intuitive, yet extensible
fluctuating hydrodynamic theory. In particular, we compare to previous
mode-mode coupling theories of the long-time tails in the velocity
autocorrelation function, as well as to theories of finite-size effects
on diffusion in periodic systems. Furthermore, in Section \ref{sub:VACF_MD}
we re-examine existing data from hard-disk molecular dynamics simulations
to find that the simple theory describes the system size dependence
of the long-time diffusion coefficient of hard disks as well, confirming
that the phenomenon we study is a generic property of particle fluids
and not an artifact of DSMC. For large system sizes, however, we find
that a more sophisticated self-consistent theory is necessary, and
make some preliminary attempts at an explicit self-consistent calculation
in both two and three dimensions, before offering concluding remarks
and discussing future research directions.

\section{\label{sec:Fluctuating-Hydrodynamics}Fluctuating Hydrodynamics}

At mesoscopic scales the hydrodynamic behavior of fluids can be described
with continuum stochastic PDEs of the Langevin type \citet{GardinerBook,vanKampen:07}.
Thermal fluctuations enter as random forcing terms in the Landau-Lifshitz
Navier-Stokes (LLNS) equations of fluctuating hydrodynamics \citet{Landau:Fluid,LLNS_Espanol}.
For a mixture of two indistinguishable fluids, neglecting viscous
heating, the compressible LLNS equations are \citet{FluctHydroNonEq_Book,Bell:09}
\begin{align}
D_{t}\rho= & -\rho\left(\grad\cdot\V v\right)\nonumber \\
\rho\left(D_{t}\V v\right)= & -\grad P+\grad\cdot\left(\eta\overline{\grad}\V v+\M{\Sigma}\right)\nonumber \\
\rho c_{v}\left(D_{t}T\right)= & -P\left(\grad\cdot\V v\right)+\grad\cdot\left(\kappa\grad T+\M{\Xi}\right)\nonumber \\
\rho\left(D_{t}c\right)= & \grad\cdot\left[\rho\chi\left(\grad c\right)+\M{\Psi}\right],\label{LLNS_primitive}\end{align}
where $D_{t}\square=\partial_{t}\square+\V v\cdot\grad\left(\square\right)$
is the advective derivative, $\overline{\grad}\V v=(\grad\V v+\grad\V v^{T})-2\left(\grad\cdot\V v\right)\M I/3$,
and the pressure is $P=\rho\left(k_{B}T/m\right)=\rho c_{T}^{2}$,
where $c_{T}$ is the isothermal speed of sound. The viscosity $\eta$,
thermal conductivity $\kappa$, and the mass diffusion coefficient
$\chi$ may in general depend on the state. The capital Greek letters
denote stochastic fluxes that are modeled as white-noise random Gaussian
tensor and vector fields, with amplitudes determined from the fluctuation-dissipation
balance principle \citet{FluctuationDissipation_Kubo},\begin{align}
\M{\Sigma}=\sqrt{2\eta k_{B}T}\,\M{\mathcal{W}},\quad & \M{\Psi}=\sqrt{2m\chi\rho\; c(1-c)}\,\widetilde{\M{\mathcal{W}}},\mbox{ and }\M{\Xi}=T\sqrt{2\kappa k_{B}}\,\breve{\M{\mathcal{W}}}\label{stoch_flux_covariance}\end{align}
where $m$ is the fluid particle mass, and $\M{\mathcal{W}}$, $\widetilde{\M{\mathcal{W}}}$
and $\breve{\M{\mathcal{W}}}$ are white-noise random Gaussian tensor
and vector fields with covariance \begin{align*}
\av{\mathcal{W}_{ij}(\V r,t)\mathcal{W}_{kl}^{\star}(\V r^{\prime},t^{\prime})} & =\left(\delta_{ik}\delta_{jl}+\delta_{il}\delta_{jk}-2\delta_{ij}\delta_{kl}/3\right)\delta(t-t^{\prime})\delta(\V r-\V r^{\prime}),\\
\av{\widetilde{\mathcal{W}}_{i}(\V r,t)\widetilde{\mathcal{W}}_{j}^{\star}(\V r^{\prime},t^{\prime})} & =\av{\breve{\mathcal{W}}_{i}(\V r,t)\breve{\mathcal{W}}_{j}^{\star}(\V r^{\prime},t^{\prime})}=\left(\delta_{ij}\right)\delta(t-t^{\prime})\delta(\V r-\V r^{\prime}),\end{align*}
where star denotes complex conjugate.

In addition to the usual Fickian contribution, the flux in the equation
(\ref{LLNS_primitive}) for $c=\bar{c}+\d c$ includes advection by
the fluctuating velocities, $\V v=\bar{\V v}+\d{\V v}=\d{\V v}$.
Ignoring density fluctuations,\begin{equation}
\partial_{t}\left(\delta c\right)+\left(\d{\V v}\right)\cdot\left(\grad\bar{c}\right)=\grad\cdot\left[\chi\grad\left(\delta c\right)-\left(\delta c\right)\left(\d{\V v}\right)\right]+\rho^{-1}\left(\grad\cdot\M{\Psi}\right).\label{eq:dc_t_nonlinear}\end{equation}
Interpreting the non-linear stochastic partial differential equation
(SPDE) (\ref{eq:dc_t_nonlinear}) requires some form of \emph{regularization}
(smoothing) of the stochastic forcing, usually approached using a
perturbative approach \citet{DiffusionRenormalization_I,DiffusionRenormalization_II,Renormalization_Mazur,ExtraDiffusion_Vailati}.
To leading order, we can approximate the advective contribution to
the average diffusive mass flux, using the solution of the \emph{linearized}
equations of fluctuating hydrodynamics, which can be given a precise
meaning \citet{DaPratoBook}. Specifically, we anticipate a relation
of the form \[
-\av{\left(\delta c\right)\left(\d{\V v}\right)}\approx-\av{\left(\delta c\right)\left(\d{\V v}\right)}_{\text{linear}}=\left(\D{\chi}\right)\grad\bar{c},\]
leading to an \emph{effective }diffusion coefficient $\chi_{\text{eff}}=\chi+\D{\chi}$
that includes an \emph{enhancement} $\D{\chi}$ due to thermal velocity
fluctuations, in addition to the \emph{bare} diffusion coefficient
$\chi$. In Appendix \ref{AppendixEstimates} we give some simple
estimates of the relative magnitude of $\D{\chi}$ in relation to
$\chi$, demonstrating that the enhancement due to velocity fluctuations
is expected to be much larger for dense liquids than for dilute gases.

\subsection{Fluctuation-Enhanced Diffusion Coefficient}

In order to analyze the stationary solution to the linearized equations
of fluctuating hydrodynamics, we will apply a Fourier transform in
all directions as done in Ref. \citet{ExtraDiffusion_Vailati}, even
though the direction of the gradient is not periodic. One can justify
this approximation by considering a periodic background concentration
field, maintained at steady state via some external potential, and
then calculate the mass flux in the vicinity of the plane $y=L_{y}/2$
as a function of the local concentration gradient in the limit of
infinite period $L_{y}$ \citet{FluctHydroNonEq_Book}.

To simplify the analysis, we can neglect density and temperature variations,
$\rho=\rho_{0}$ and $T=T_{0}$, to obtain the \emph{isothermal incompressible}
approximation,\begin{align}
\partial_{t}\V v= & \M{\mathcal{P}}\left[-\V v\cdot\grad\V v+\nu\grad^{2}\V v+\rho^{-1}\left(\grad\cdot\M{\Sigma}\right)\right]\label{eq:LLNS_incomp_v}\\
\partial_{t}c= & -\V v\cdot\grad c+\chi\grad^{2}c+\rho^{-1}\left(\grad\cdot\M{\Psi}\right).\label{eq:LLNS_incomp_c}\end{align}
where $\nu=\eta/\rho$, and $\V v\cdot\grad c=\grad\cdot\left(c\V v\right)$
and $\V v\cdot\grad\V v=\grad\cdot\left(\V v\V v^{T}\right)$ because
of incompressibility, $\grad\cdot\V v=0$. Here $\M{\mathcal{P}}$
is the orthogonal projection onto the space of divergence-free velocity
fields, $\widehat{\M{\mathcal{P}}}=\M I-k^{-2}(\V k\V k^{\star})$
in Fourier space (denoted with a hat).

The linearized form of (\ref{eq:LLNS_incomp_v},\ref{eq:LLNS_incomp_c})
in the Fourier domain is a collection of stochastic differential equations,
one system of linear additive-noise equations per wavenumber, of the
form \begin{equation}
d\left[\begin{array}{c}
\widehat{\delta c}\\
\widehat{\delta\V v}\end{array}\right]=-\left[\begin{array}{cc}
\nu\, k^{2}\widehat{\M{\mathcal{P}}} & \V 0\\
\V g_{c} & \chi k^{2}\end{array}\right]\left[\begin{array}{c}
\widehat{\delta c}\\
\widehat{\delta\V v}\end{array}\right]dt+\left[\begin{array}{cc}
2\rho^{-1}\nu k_{B}T\, k^{2}\widehat{\M{\mathcal{P}}} & \V 0\\
\V 0 & 2\rho^{-1}\chi\, mc(1-c)\, k^{2}\end{array}\right]^{1/2}d\M{\mathcal{B}},\label{eq:cv_t_hat}\end{equation}
where $\V{\mathcal{B}}$ is a collection of independent Wiener processes.
At steady state the correlations between the Gaussian fluctuations
are described by the matrix of static structure factors (covariance
matrix)\[
\M{\mathcal{S}}=\left[\begin{array}{cc}
\av{(\widehat{\delta\V v})(\widehat{\d{\V v}})^{\star}} & \av{(\widehat{\d{\V v}})(\widehat{\d c})^{\star}}\\
\av{(\widehat{\delta c})(\widehat{\d{\V v}})^{\star}} & \av{(\widehat{\delta c})(\widehat{\d c})^{\star}}\end{array}\right].\]
The static structure factor matrix consists of a \emph{short-ranged}
equilibrium contribution and a \emph{long-range} non-equilibrium contribution,
\[
\M{\mathcal{S}}=\left[\begin{array}{cc}
\rho^{-1}k_{B}T\,\widehat{\M{\mathcal{P}}} & 0\\
0 & m\rho^{-1}\, c(1-c)\end{array}\right]+\nabla\bar{c}\left[\begin{array}{cc}
0 & \D{\M{\mathcal{S}}}_{c,\V v}^{\star}\\
\D{\M{\mathcal{S}}}_{c,\V v} & \left(\D S_{c,c}\right)\nabla\bar{c}\end{array}\right].\]
The explicit form of $\M{\mathcal{S}}$ can be obtained as the solution
of a linear system derived from (\ref{eq:cv_t_hat}) using the stationarity
condition $d\M{\mathcal{S}}=\M 0$ \citet{LLNS_S_k}. The concentration
fluctuations are enhanced as the square of the applied gradient \citet{ExtraDiffusion_Vailati},\begin{equation}
\av{(\widehat{\delta c})(\widehat{\d c})^{\star}}_{\text{neq}}=\left(\D S_{c,c}\right)\left(\nabla\bar{c}\right)^{2}=\frac{k_{B}T}{\rho\chi(\nu+\chi)k^{4}}\left(\sin^{2}\theta\right)\,\left(\nabla\bar{c}\right)^{2},\label{eq:S_c_c}\end{equation}
while the correlation between the concentration fluctuations and the
fluctuations of velocity parallel to the concentration gradient are
linear in the applied gradient \citet{ExtraDiffusion_Vailati}, \begin{equation}
\av{(\widehat{\delta c})(\widehat{\d v}_{\Vert}^{\star})}=\mathcal{S}_{c,v_{\Vert}}=\left(\D{\mathcal{S}}_{c,v_{\Vert}}\right)\nabla\bar{c}=-\frac{k_{B}T}{\rho(\nu+\chi)k^{2}}\left(\sin^{2}\theta\right)\,\nabla\bar{c}.\label{eq:S_c_vy}\end{equation}
where $\theta$ is the angle between $\V k$ and $\grad\bar{c}$,
$\sin^{2}\theta=k_{\perp}^{2}/k^{2}$. The power-law divergence for
small $k$ indicates long ranged correlations between $\delta c$
and $\d{\V v}$ and is the cause of both the giant fluctuation phenomenon
and the diffusion enhancement. As seen from (\ref{eq:av_j1}), the
actual correlation that determines the diffusion enhancement is $\mathcal{S}_{\rho_{1},v_{\Vert}^{(1)}}=\av{(\widehat{\delta c})(\widehat{\d v}_{\Vert}^{(1)})^{\star}}$,
which is approximated as $\mathcal{S}_{\rho_{1},v_{\Vert}^{(1)}}\approx\bar{\rho}\mathcal{S}_{c,v_{\Vert}}$
in (\ref{eq:LLNS_incomp_v},\ref{eq:LLNS_incomp_c}); this approximation
is discussed and justified in Section \ref{sub:Static-Factor}.

The mass flux due to advection by the fluctuating velocities can be
approximated as\begin{equation}
\av{\delta c\left(\V r,t\right)\d{\V v}\left(\V r,t\right)}=\left(2\pi\right)^{-6}\int d\V k\, d\V k^{\prime}\,\av{\widehat{\delta c}\left(\V k,t\right)\widehat{\delta\V v}^{\star}\left(\V k^{\prime},t\right)}e^{i\left(\V k-\V k^{\prime}\right)\cdot\V r}=\left(2\pi\right)^{-3}\int_{\V k}\mathcal{S}_{c,\V v}\left(\V k\right)\, d\V k,\label{eq:dcdv_realspace}\end{equation}
which together with (\ref{eq:S_c_vy}) gives an estimate of the diffusion
enhancement \citet{ExtraDiffusion_Vailati},\begin{equation}
\D{\chi}=-\left(2\pi\right)^{-3}\int_{\V k}\D{\mathcal{S}}_{c,v_{\Vert}}\left(\V k\right)\, d\V k=\frac{k_{B}T}{(2\pi)^{3}\rho\left(\chi+\nu\right)}\;\int_{\V k}\left(\sin^{2}\theta\right)k^{-2}\, d\V k.\label{eq:dchi_k_int}\end{equation}
Because of the $k^{-2}$-like behavior, the integral over all $\V k$
above diverges unless one imposes a lower bound, $k_{\min}\sim2\pi/L$
in the absence of gravity, \emph{and} a phenomenological cutoff $k_{\max}\sim\pi/L_{\text{mol}}$
\citet{ExtraDiffusion_Vailati} for the upper bound, where $L_{\text{mol}}$
is an ad-hoc {}``molecular'' length scale. Importantly, the fluctuation
enhancement $\D{\chi}$ depends on the system size $L$ because of
the small wavenumber cutoff.

\subsubsection{\label{sub:Two-Dimensions}Two Dimensions}

For a quasi two-dimensional system, $L_{z}\ll L_{x}\ll L_{y}$, we
can replace the integral over $k_{z}$ with $2\pi/L_{z}$ and integrate
over all $k_{y}$. This leads to an average total mass flux that grows
logarithmically with the system width $L_{x}$ for a fixed height
$L_{y}$, \begin{equation}
\D{\chi}\left(L_{x}\right)\approx\frac{k_{B}T}{(2\pi)^{2}\rho\left(\chi+\nu\right)L_{z}}\,\int_{\abs{k_{x}}\geq2\pi/L_{x}}^{\abs{k_{x}}\leq\pi/L_{\text{mol}}}\frac{k_{x}^{2}}{\left(k_{x}^{2}+k_{y}^{2}\right)^{2}}\, dk_{y}dk_{x}=\frac{k_{B}T}{4\pi\rho(\chi+\nu)L_{z}}\,\ln\frac{L_{x}}{2L_{\text{mol}}}.\label{eq:j_lnLx}\end{equation}
When the system width (perpendicular to the gradient) becomes comparable
to the height (parallel to the gradient), boundaries will intervene
and for $L_{x}\gg L_{y}$ the effective diffusion coefficient must
become a constant, which is predicted to be a logarithmically-growing
function of $L_{y}$ in two dimensions.

It is important to emphasize here that the chosen value of $L_{\text{mol}}$
is arbitrary. The hydrodynamic theory models the effective diffusion
coefficient as the sum of the {}``bare'' diffusion coefficient $\chi$
and the {}``enhancement'' $\D{\chi}$, but the two cannot be separated
because every measurement must be performed for some \emph{finite}
$L_{x}$. One can thus simply define $\chi$ to be the value of the
measured diffusion coefficient for some reference width $L_{0}>2L_{\text{mol}}$,
and predict that for $L_{x}\geq L_{0}$,\begin{equation}
\chi_{\text{eff}}^{(2D)}\approx\chi+\frac{k_{B}T}{4\pi\rho(\chi+\nu)L_{z}}\,\ln\frac{L_{x}}{L_{0}}.\label{eq:chi_eff_2D}\end{equation}
For this prediction to be accurate, however, $L_{0}$ ought to be
chosen to not be too large, so that the enhancement of the diffusion
relative to the \textquotedbl{}molecular\textquotedbl{} contributions
is small and simple quasi-linearized theory applies, but also not
too small so that fluctuating hydrodynamics applies.

Because we are explicitly concerned with the effect of a \emph{finite}
width $L_{x}$, the integral over $k_{x}$ should be replaced by a
discrete sum over the wavenumbers consistent with periodicity, $k_{x}=\kappa_{x}\cdot2\pi/L_{x}$,
where $\kappa_{x}\in\Set Z$. If one calculates the \emph{difference}
between a system of width $2L_{x}$ and a system of width $L_{x}$,
then it is easily seen that the integral over $k_{x}$ in Eq. (\ref{eq:j_lnLx})
ought to be replaced with the following sum over $\kappa_{x}$,\[
\left(2\pi\right)^{-1}\,\left[\int_{\abs{k_{x}}\geq\pi/L_{x}}f(k_{x})\, dk_{x}-\int_{\abs{k_{x}}\geq2\pi/L_{x}}f(k_{x})\, dk_{x}\right]\;\longleftrightarrow\; L_{x}^{-1}\,\sum_{\kappa_{x}\neq0}\left[f\left(2\kappa_{x}-1\right)-f\left(2\kappa_{x}\right)\right].\]
Even though $f\left(\kappa_{x}\right)\sim k_{x}^{-1}$ in (\ref{eq:j_lnLx})
is not integrable, the difference in the square bracket above goes
like $\kappa_{x}^{-2}$ and the sum can be done explicitly, giving
exactly the same answer as the integral estimate, \[
\chi_{\text{eff}}^{(2D)}\left(2L_{x}\right)-\chi_{\text{eff}}^{(2D)}\left(L_{x}\right)=\frac{k_{B}T}{4\pi\rho(\chi+\nu)L_{z}}\,\ln\,2.\]

\subsubsection{Three Dimensions}

Now we consider a system where $L_{x}=L_{z}=L\ll L_{y}$, and study
how the effective diffusion coefficient changes with $L$. In three
dimensions, the relative contribution from large wavenumbers, i.e.,
small scales, is larger than in two dimensions. We can use the integral
approximation to examine the asymptotic behavior for large $L$,\[
\D{\chi}\approx\frac{k_{B}T}{(2\pi)^{3}\rho\left(\chi+\nu\right)}\,\int_{\abs{k_{x/z}}\geq2\pi/L}^{\abs{k_{x/z}}\leq\pi/L_{\text{mol}}}\frac{k_{x}^{2}+k_{z}^{2}}{\left(k_{x}^{2}+k_{y}^{2}+k_{z}^{2}\right)^{2}}\, dk_{y}dk_{x}dk_{z}=\frac{\ln\left(1+\sqrt{2}\right)\, k_{B}T}{\pi\rho(\chi+\nu)}\left(\frac{1}{2L_{\text{mol}}}-\frac{1}{L}\right).\]
We see that in three dimensions $\chi_{\text{eff}}$ converges as
$L\rightarrow\infty$ to the \emph{macroscopic} diffusion coefficient,
but for a finite system the effective diffusion coefficient is reduced
by an amount $\sim L^{-1}$ due to the truncation of the velocity
fluctuations by the confining walls,\begin{equation}
\chi_{\text{eff}}^{(3D)}\approx\chi+\frac{\alpha\, k_{B}T}{\rho(\chi+\nu)}\left(\frac{1}{L_{0}}-\frac{1}{L}\right).\label{eq:chi_eff_3D}\end{equation}
Calculating the exact value of $\alpha$ requires performing a sum
over $\kappa_{x}$ and $\kappa_{z}$ instead of integrals over $k_{x}$
and $k_{z}$, as we have done numerically. The numerical results suggest
that, as in two dimensions, the difference in $\chi_{\text{eff}}^{(3D)}$
between two systems attains a finite value as $L_{\text{mol}}\rightarrow0$,
justifying (\ref{eq:chi_eff_3D}) for $\left(L_{0},L\right)\gg L_{\text{mol}}$.

\section{\label{sec:Particle-Simulations}Particle Simulations}

This sections verifies the predictions of fluctuating hydrodynamics
by particle simulations. Here we employ the Direct Simulation Monte
Carlo (DSMC) particle algorithm \citet{DSMC_Bird,DSMCReview_Garcia},
in which deterministic interactions between the particles are replaced
with stochastic collisions exchanging momentum and energy between
nearby particles. The collision rules ensure local energy and momentum
conservation and a thermodynamically-consistent fluctuation spectrum
\citet{GranularFluctuations_DSMC,SHSD}. Previous careful measurements
of transport coefficients in DSMC using nonequilibrium methods have
been limited to quasi one-dimensional simulations, in which there
is only one collision cell along the dimensions perpendicular to the
gradient \citet{DSMCConductivity_Gallis}. The effect we are exploring
here does not appear in one dimension as it arises because of the
presence of vortical modes in the fluctuating velocities.

We have performed DSMC calculations for an ideal hard-sphere gas with
molecular diameter $\sigma=1$ and molecular mass $m=1$, at an equilibrium
density of $\rho_{0}=0.06$, with the temperature kept at $k_{B}T_{0}=T_{0}=1$
via thermal collisions with the top and bottom walls. A uniform concentration
gradient along the vertical ($y$) direction is implemented by randomly
selecting the species of particles to be one with probability $c_{T/B}$
when they collide with the top/bottom wall. Each DSMC particle represents
a single hard sphere so the mean free path is $\lambda=3.75$ and
the mean free collision time is $\tau=2.35$. The DSMC time step was
chosen to be $\D t=\tau/2$, and the collision cell size is either
$\D x_{c}=\lambda$ or $\D x_{c}=2\lambda$.

The DSMC algorithm simulates a dilute gas, for which the enhancement
of diffusion is weaker than for dense fluids (see Appendix \ref{AppendixEstimates}).
Nevertheless, the computational efficiency of DSMC makes it preferable
to molecular dynamics for this study. The DSMC method employed here
uses a grid of collision cells, thus introducing discretization artifacts
into the particle dynamics. While it is possible to eliminate these
grid effects entirely \citet{SHSD_PRL,SHSD}, the associated increase
in computational cost and the difficulty of parallelization would
make some of the large-scale particle simulations presented here infeasible.
Furthermore, decreasing the density in order to increase the mean
free path and reduce the grid effects would make the relative size
of the effect we are trying to observe too small compared to statistical
errors. We have verified that quantitatively identical results are
obtained for two different choices of DSMC collision cells, $\D x_{c}=\lambda$
or $\D x_{c}=2\lambda$, once the discretization correction to Chapman-Enskog
kinetic theory for the transport coefficients is taken into account
\citet{DSMC_CellSizeError,DSMC_TimeStepError,DSMC_TimeStepError2}.

In addition to the DSMC collision cells, which determine the microscopic
dynamics of our particle simulations, obtaining hydrodynamic quantities
such as velocity requires using a grid of $N_{x}\times N_{y}\times N_{z}$
\emph{sampling} or \emph{hydrodynamic cells}, each of volume $\D V=\left|\D{\mathcal{V}}\right|=\D x\,\D y\,\D z$.
The sampling of hydrodynamic quantities is performed every $5$ DSMC
time steps, at a snapshot time that is \emph{randomly} chosen. Sampling
at random time intervals ensures that there is no measurement bias
due to the lack of time invariance in the particle dynamics, and gives
similar results as sampling at the mid point of each time step \citet{DSMCConductivity_Gallis}.
At each snapshot, we obtain the \emph{instantaneous} mass $m_{\D{\mathcal{V}}}=\left(\D V\right)\rho_{\D{\mathcal{V}}}$
and momentum $\V p_{\D{\mathcal{V}}}=\left(\D V\right)\V j_{\D{\mathcal{V}}}$
in each sampling cells by adding the contributions from all particles
inside the given sampling cell. We can do this sampling taking into
account either all of the particles, $\rho_{\D{\mathcal{V}}}$ and
$\V j_{\D{\mathcal{V}}}$, or just particles of the first species,
which we indicate by a species subscript or parenthesis superscript,
$\rho_{\D{\mathcal{V}}}^{(1)}$ and $\V j_{\D{\mathcal{V}}}^{(1)}$.
For each sampling cell, we obtain an \emph{instantaneous} velocity
$\V v_{\D{\mathcal{V}}}=\V j_{\D{\mathcal{V}}}/\rho_{\D{\mathcal{V}}}$
(similarly for $\V v_{\D{\mathcal{V}}}^{(1)}$) and mass concentration
$c=\rho_{\D{\mathcal{V}}}^{(1)}/\rho_{\D{\mathcal{V}}}$. We obtain
discrete static structure factors (spectral correlations) from time
averages of products of discrete Fourier transforms (DFTs) of the
instantaneous variables. For comparison between the particle simulations
and the theory we use a reference length $L_{0}=16\lambda$.

\subsection{\label{sub:Static-Factor}Static Structure Factor}

In order to compare the prediction (\ref{eq:S_c_vy}) to results from
particle simulations, we need to convert the \emph{continuum} static
structure factor $\mathcal{S}_{c,v_{\Vert}}(\V k)$ into a \emph{discrete}
structure factor $S_{c,v_{\Vert}}(\V{\kappa})$ for finite-volume
averages of the continuum fields. Here the set of $N_{x}\times N_{y}\times N_{z}$
wavenumbers $\V{\kappa}\in\Set Z^{3}$ indexes the discrete set of
wavevectors compatible with periodicity, $\V k\left(\V{\kappa}\right)=2\pi\left(\kappa_{x}L_{x}^{-1},\kappa_{y}L_{y}^{-1},\kappa_{z}L_{z}^{-1}\right)$.
A relatively straightforward calculation shows that\textcolor{red}{{}
}\begin{equation}
S_{c,v_{\Vert}}(\V{\kappa})=\sum_{\V{\kappa}^{\prime}}F_{\D{\mathcal{V}}}\left[\V k\left(\V{\kappa}^{\prime}\right)\right]\,\mathcal{S}_{c,v_{\Vert}}\left[\V k\left(\V{\kappa}^{\prime}\right)\right],\label{eq:S_k_discrete}\end{equation}
where the sum goes over all resonance modes, $\V{\kappa}^{\prime}=\left(\kappa_{x}+N_{x}\D{\kappa}_{x},\,\kappa_{y}+N_{y}\D{\kappa}_{y},\,\kappa_{z}+N_{z}\D{\kappa}_{z}\right)$
for all $\D{\V{\kappa}}\in\Set Z^{3}$, and $F_{\D{\mathcal{V}}}(\V k)=F_{x}\left(k_{x}\right)F_{y}\left(k_{y}\right)F_{z}\left(k_{z}\right)$
is a product of low-pass filters of the form\begin{equation}
F_{x}(k_{x})=\mbox{sinc}^{2}\left(k_{x}\D x/2\right),\label{eq:F_x_filt}\end{equation}
where $\mbox{sinc}(x)=\sin(x)/x$. The sum in (\ref{eq:S_k_discrete})
can easily be evaluated numerically because the terms decay rapidly
{[}c.f. (\ref{eq:S_c_vy}){]}.

\begin{figure}[h]
\begin{centering}
\includegraphics[width=0.7\columnwidth]{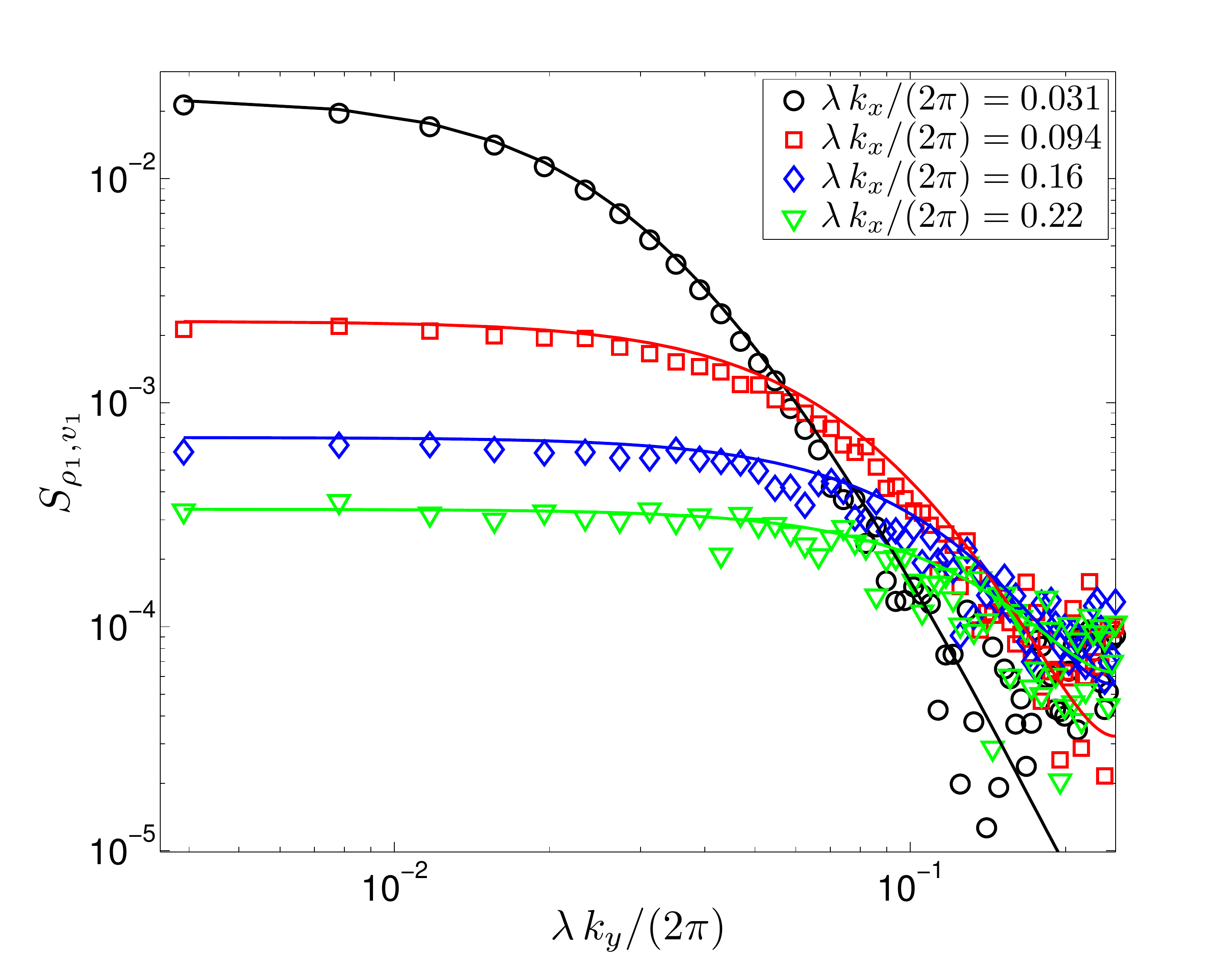}
\par\end{centering}

\begin{centering}
\includegraphics[width=0.7\columnwidth]{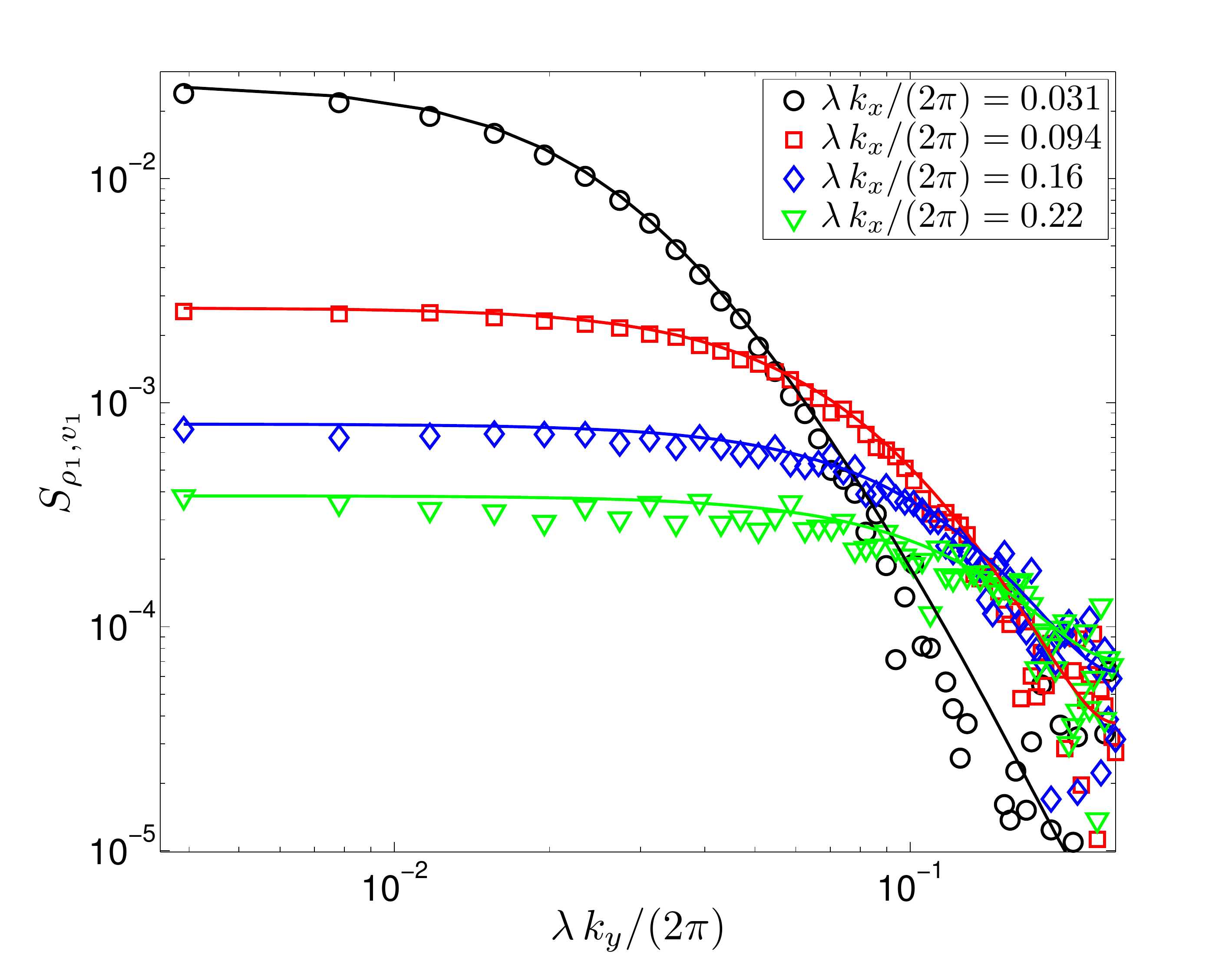}
\par\end{centering}

\caption{\label{fig:DSMC_S_rho1v1}(\emph{Top}) Discrete structure factor $\mathcal{S}_{\rho_{1},v_{\Vert}^{(1)}}$
from quasi two-dimensional DSMC runs with $L_{x}=64\lambda$, $L_{y}=512\lambda$
and $L_{z}=2\lambda$, for several wavenumbers $k_{x}=\kappa_{x}\cdot2\pi/L_{x}$
(circles $\kappa_{x}=2$, squares $\kappa_{x}=6$, diamonds $\kappa_{x}=10$,
triangles $\kappa_{x}=14$), compared to $\bar{\rho}S_{c,v_{\Vert}}$
(solid lines of the same color) as predicted by the linearized periodic
theory (\ref{eq:S_c_vy},\ref{eq:S_k_discrete}). The sides of the
DSMC collision cells are $\D x_{c}=\D y_{c}=2\lambda=7.5$. Note that
for a fixed $k_{x}$ we expect the structure factor to decay as $k_{y}^{-4}$.
(\emph{Bottom}) All the parameters, including the system and sampling
cell size, are as for the system in the top panel, but now the DSMC
cells are twice smaller, $\D x_{c}=\D y_{c}=\lambda=3.75$. Note that
this change of the DSMC cell size changes the kinetic theory prediction
for viscosity by more than 20\% \citet{DSMC_CellSizeError}.}

\end{figure}

\begin{figure}[h]
\begin{centering}
\includegraphics[width=0.7\columnwidth]{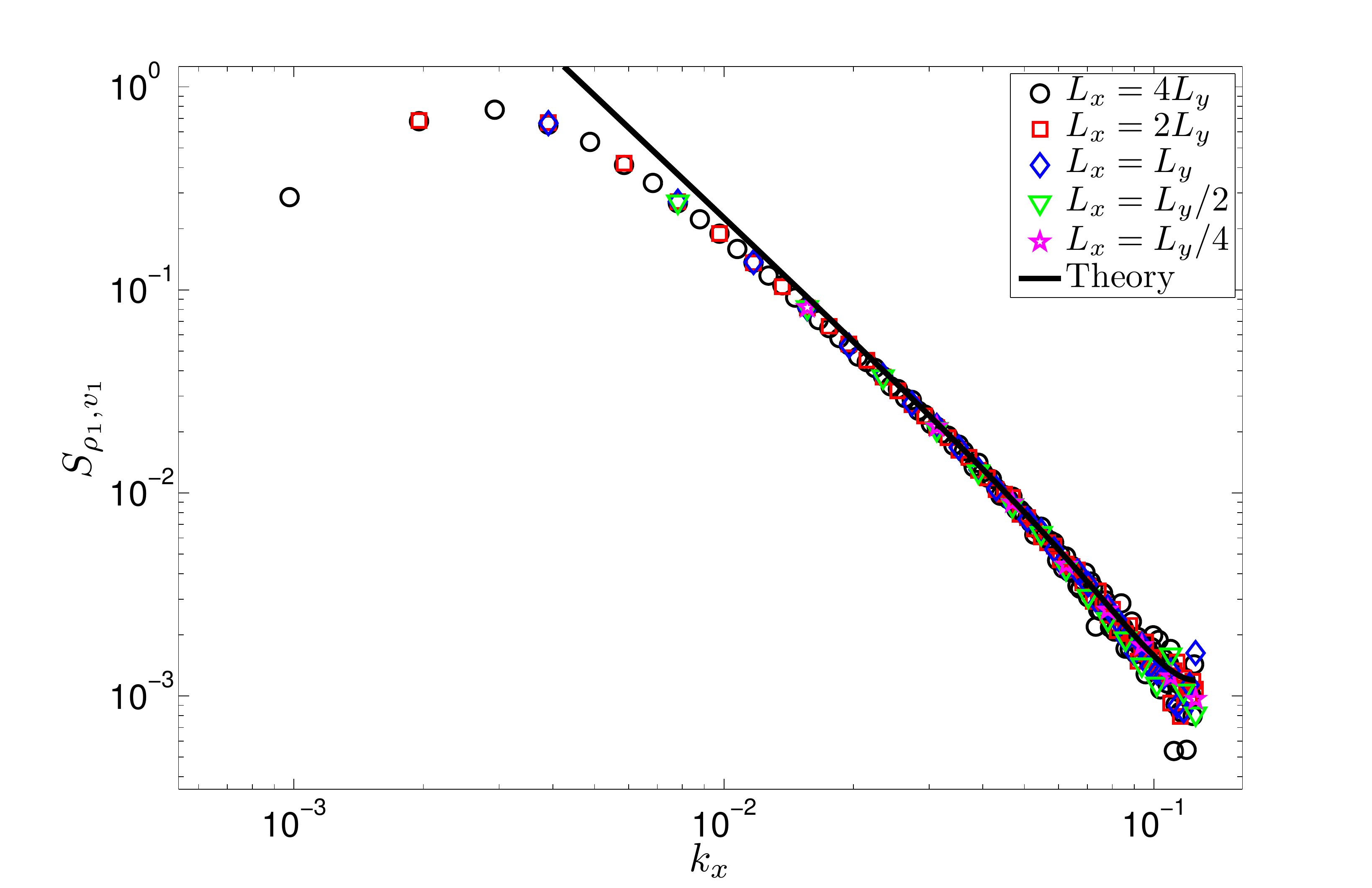}
\par\end{centering}

\caption{\label{fig:DSMC_S_rho1v1_perp}Discrete structure factor $\mathcal{S}_{\rho_{1},v_{\Vert}^{(1)}}$
from quasi two-dimensional DSMC runs with $L_{y}=512\lambda$ and
$L_{z}=\D x_{c}=\D y_{c}=2\lambda$, for wavevectors perpendicular
to the gradient ($k_{y}=0$). Results for systems of different width
$L_{x}$ are shown with symbols (see legend). For comparison, the
theoretical prediction for $\bar{\rho}S_{c,v_{\Vert}}$ for an infinite
periodic system (\ref{eq:S_c_vy},\ref{eq:S_k_discrete}) is shown
with a line. Deviations from the predicted $k_{x}^{-2}$ power-law
divergence are clear for $k_{x}\lesssim2\pi/L_{y}\approx3\cdot10^{-3}$
due to the influence of the top and bottom walls.}

\end{figure}

In Fig. \ref{fig:DSMC_S_rho1v1} we compare the theoretical prediction
for $\bar{\rho}S_{c,v_{\Vert}}(\V{\kappa})$ to results from particle
simulations for the discrete structure factor \[
\mathcal{S}_{\rho_{1},v_{\Vert}^{(1)}}(\V{\kappa})=\av{\left(\widehat{\delta\rho_{1}}\right)\left(\widehat{\d v}_{\Vert}^{(1)}\right)^{\star}},\]
for two different sizes of the DSMC collision cells. The fact that
there is little difference between the two panels in the figure verifies
that the details of the microscopic collision dynamics do not affect
the mesoscopic hydrodynamic behavior. In Fig. \ref{fig:DSMC_S_rho1v1_perp}
we plot the discrete structure factor from the particle simulations
for wavevectors perpendicular to the gradient (i.e., $k_{y}=0$),
for systems of different width $L_{x}$.

It is expected that compressibility effects would affect $\mathcal{S}_{\rho_{1},v_{\Vert}^{(1)}}$.
Indeed, this is what we observe in simulations, however, as Fig. \ref{fig:DSMC_S_rho1v1}
demonstrates, the incompressible isothermal theory for $\bar{\rho}S_{c,v_{\Vert}}$
is in very good agreement with particle data for $\mathcal{S}_{\rho_{1},v_{\Vert}^{(1)}}$.
Good agreement between the simulation data and the simple theory is
also seen in Fig. \ref{fig:DSMC_S_rho1v1_perp}, except for $k_{x}$
comparable to $2\pi/L_{y}$. As expected, for the smallest wavenumbers
the top and bottom walls intervene and the actual correlation is smaller
than the predicted $k_{x}^{-2}$ divergence.

In order to construct a theoretical prediction for $\mathcal{S}_{\rho_{1},v_{\Vert}^{(1)}}$,
one must not only include the effects of compressibility but also
replace the {}``one-fluid'' approximation (\ref{LLNS_primitive})
with a corresponding {}``two-fluid'' compressible hydrodynamic theory
\citet{BinaryMixKineticTheory,MultiFluidKineticTheory}. This can
be seen by noting that the fluctuating equations (\ref{LLNS_primitive})
assume that relation (\ref{eq:v12_bar}) applies to the fluctuating
$\V v_{12}$ and $c$ instead of their means. Such an assumption leads
to unphysical bias of order $\av{\left(\d c\right)^{2}}$ in the mean
inter-species velocity $\av{\V v_{12}}$ because of the nonlinearity
in the denominator $c(1-c)$. In fact, the fluctuations $\d{\V v_{12}}$
and $\d c$ should be uncorrelated, as seen from a two-fluid fluctuating
theory. Here we use the incompressible isothermal approximation for
$\bar{\rho}S_{c,v_{\Vert}}$ as a proxy for $\mathcal{S}_{\rho_{1},v_{\Vert}^{(1)}}$
in order to construct theoretical predictions for the diffusion enhancement.

\subsection{\label{sub:DSMC_Results}Fluctuation-Enhanced Diffusion Coefficient}

As we already explained, instantaneous hydrodynamic quantities, denoted
with a subscript $\D{\mathcal{V}}$, are sampled from the particle
data by taking snapshots of the particle state using a grid of sampling
cells of volume $\D V$. The ensemble average of a given quantity,
which we will denote with angle brackets, is obtained by averaging
over many snapshots once a steady state is reached, and additional
averaging can be performed over all sampling cells with the same $y$
position since the steady state averages cannot depend on $x$ and
$z$. We estimate the \emph{macroscopic} mean mass density $\bar{\rho}$,
partial density $\bar{\rho}_{1}$, partial momentum density $\bar{\V j}_{1}$,
partial velocity $\bar{\V v}_{1}$ and concentration $\bar{c}$ as
\[
\bar{\rho}=\rho_{0}=\av{\rho_{\D{\mathcal{V}}}},\quad\bar{\rho}_{1}=\av{\rho_{\D{\mathcal{V}}}^{(1)}}=\bar{c}\bar{\rho}\mbox{ and }\bar{\V j}_{1}=\V j_{0}^{(1)}=\av{\V j_{\D{\mathcal{V}}}^{(1)}}=\bar{\rho}_{1}\bar{\V v}_{1}.\]
We also define the \emph{mesoscopic} velocity and concentrations to
be the ensemble averages of the instantaneous values,\[
\V v_{0}^{(1)}=\av{\V v_{\D{\mathcal{V}}}^{(1)}}\mbox{ and }c_{0}=\av{c_{\D{\mathcal{V}}}},\]
where the subscript zero will be used to simplify the cumbersome notation.
It is important to point out that for non-conserved quantities such
as $\V v$ and $c$ the mesoscopic mean can be different from the
macroscopic mean due to fluctuations \citet{Tysanner:04,UnbiasedEstimates_Garcia},
$\bar{\V v}_{1}\neq\V v_{0}^{(1)}$ and $\bar{c}\neq c_{0}$. For
conserved quantities (e.g., $\bar{\V j}_{1}$ and $\V j_{0}^{(1)}$),
however, the mesoscopic and macroscopic ensemble means are equal and
in fact independent of $\D x$ and $\D z$ (but not necessarily $\D y$).

In particle simulations, we calculate the \emph{effective} diffusion
coefficient $\chi_{\text{eff}}$ from the momentum density of one
of the species along the vertical direction,\begin{equation}
\bar{j}_{\parallel}^{(1)}=\rho_{0}\chi_{\text{eff}}\,\frac{\bar{c}_{T}-\bar{c}_{B}}{L_{y}-\D y}\approx\rho_{0}\chi_{\text{eff}}\,\nabla\bar{c},\label{eq:eff_definition}\end{equation}
where we measure $\bar{c}_{T}$ and $\bar{c}_{B}$ in the top and
bottom layer of sampling cells (whose centers are a distance $L_{y}-\D y$
from each other) to empirically account for the small concentration
slip in DSMC (about $0.5\%$ with these parameters). Numerical experiments
have verified that $\bar{j}_{\parallel}^{(1)}$ matches the flux obtained
from counting the average number of color flips at the top or bottom
walls. Furthermore, the results verify that $\chi_{\text{eff}}$ is
essentially independent of the magnitude of the concentration gradient,
and that the change in the effective gradient $\nabla\bar{c}$ as
$L_{x}$ or $L_{z}$ is changed, keeping $L_{y}$ fixed, is much smaller
than the change in $\chi_{\text{eff}}$.

\begin{figure}[h]
\begin{centering}
\includegraphics[width=0.75\columnwidth]{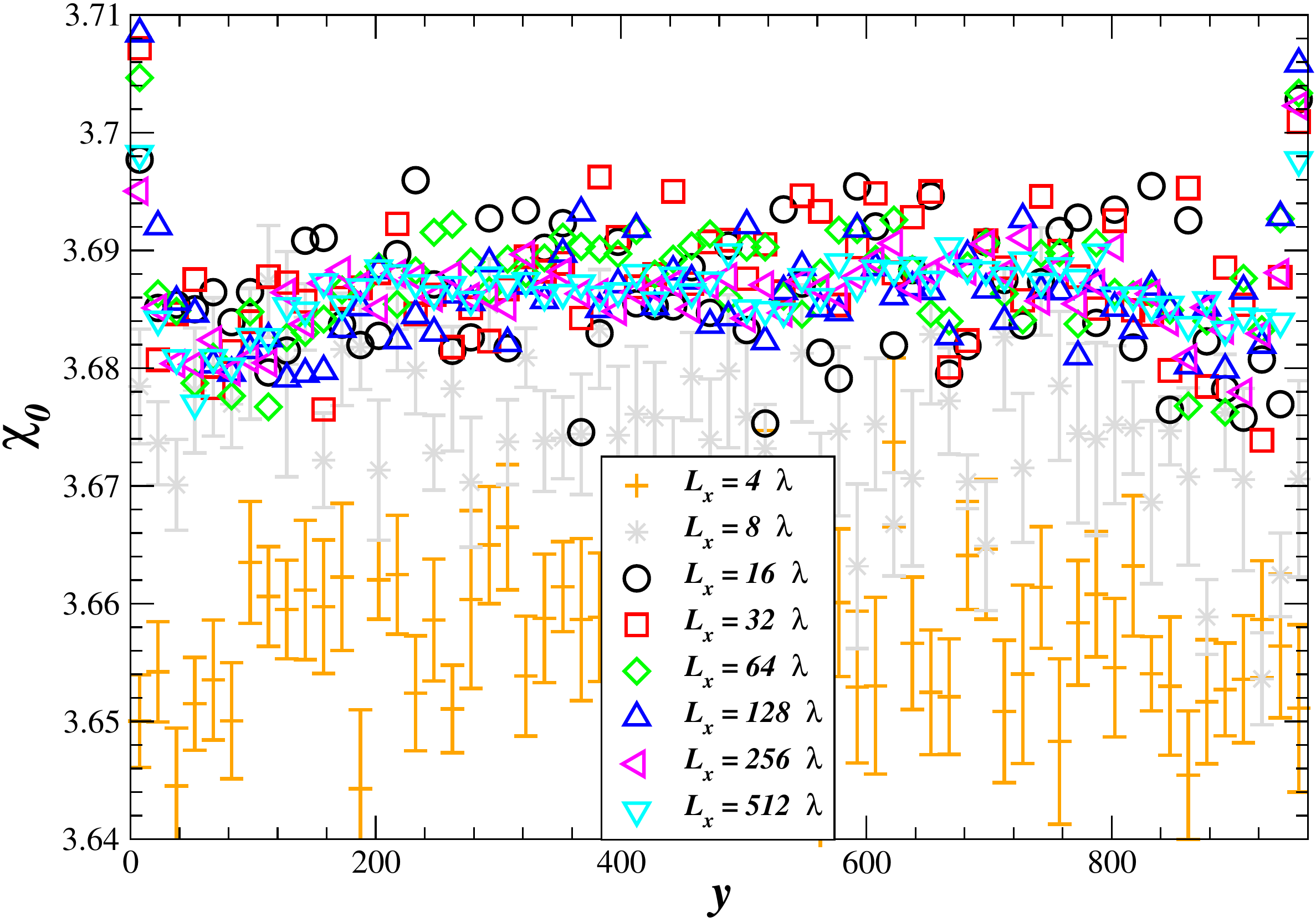}
\par\end{centering}

\caption{\label{fig:j1det_vs_y}The renormalized diffusion coefficient $\chi_{0}$,
as defined by Eq. (\ref{eq:bare_definition}), as a function of the
$y$ position of the sampling cell for DSMC systems with several system
widths $L_{x}$. The estimate of $\chi_{0}$ shown in Fig. \ref{fig:j1_vs_Nx}
is simply the average over all sampling cells further than $10\D y$
away from the top and bottom walls.}

\end{figure}

The traditional definition of a {}``renormalized'' diffusion coefficient
\citet{DiffusionRenormalization_I,DiffusionRenormalization_II} as
the macroscopic limit of $\chi_{\text{eff}}$, only works in three
dimensions and is not very useful for confined systems. Instead, for
each sampling cell, we define a \emph{locally renormalized} diffusion
coefficient $\chi_{0}$ via\begin{equation}
\rho_{0}^{(1)}\V v_{0}^{(1)}=\av{\rho_{\D{\mathcal{V}}}^{(1)}}\av{\V v_{\D{\mathcal{V}}}^{(1)}}=\bar{\rho}\chi_{0}\left(\grad\bar{c}\right),\label{eq:bare_definition}\end{equation}
where we have accounted for the fact that the macroscopic concentration
gradient $d\bar{c}/dy$ may depend on $y$. In fact, such a dependence
is observed in the particle simulations, and we have approximated
the local concentration gradient $d\bar{c}/dy$ by a numerical derivative
of a polynomial fit of degree five to $\bar{c}(y)$. Figure \ref{fig:j1det_vs_y}
shows that the empirical $\chi_{0}$ is independent of $y$, except
for a boundary layer close to the top and bottom walls. This is an
important finding, since (\ref{eq:bare_definition}) is a constitutive
model that is assumed to hold not just at the macroscale but also
at the mesoscale, notably, $\chi_{0}$ is an input parameter for fluctuating
hydrodynamics finite-volume solvers \citet{LLNS_S_k}.

\begin{figure*}[h]
\begin{centering}
\includegraphics[width=0.7\textwidth]{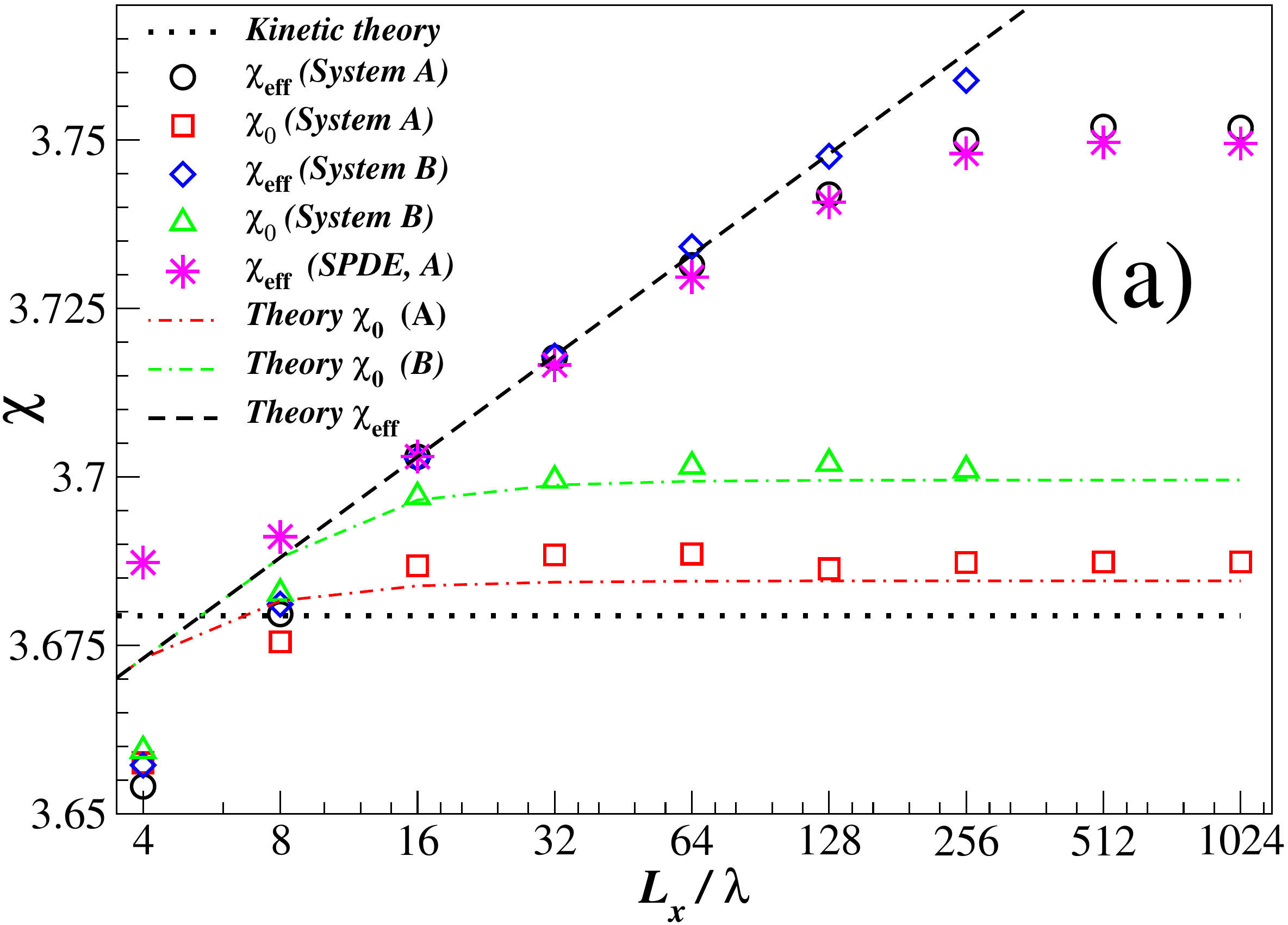}
\par\end{centering}

\begin{centering}
\includegraphics[width=0.7\textwidth]{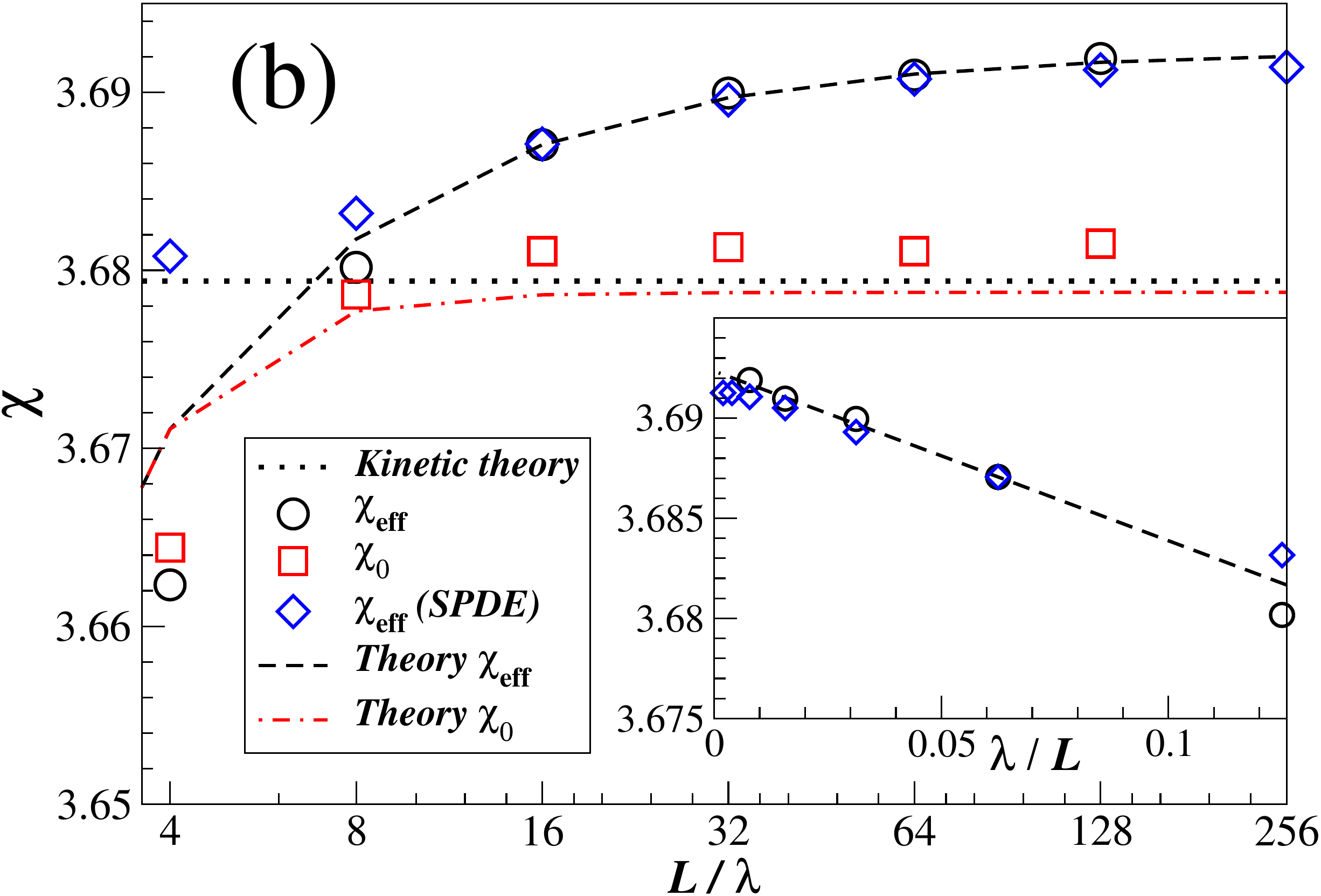}
\par\end{centering}

\caption{\label{fig:j1_vs_Nx}(\emph{Panel a, top}) The effective $\chi_{\text{eff}}$
and the renormalized $\chi_{0}$ diffusion coefficients as a function
of the width of the system $L_{x}$ in two dimensions. Numerical results
for System A (DSMC and SPDE) and System B (DSMC) are shown with symbols
(see legend). The error bars for all of the numerical data are comparable
or smaller than the size of the symbols. The theoretical predictions
(\ref{eq:chi_definitions}) are evaluated numerically and shown with
lines. The bare diffusion in the theory and SPDE calculations is adjusted
so that for $L_{x}=16\lambda$ the effective diffusion is the same
as that measured in the particle simulations. (\emph{Panel b, bottom})
Same as the top panel but in three dimensions. The inset highlights
the $L^{-1}$ behavior.}

\end{figure*}

Figure \ref{fig:j1_vs_Nx}a shows how the effective $\chi_{\text{eff}}$
and renormalized $\chi_{0}$ diffusion coefficients change as the
width of the system $L_{x}$ is increased while keeping the height
$L_{y}$ fixed for two different quasi two-dimensional DSMC systems.
For System A, the DSMC collision cells are cubes of side $\D x_{c}=7.5=2\lambda$,
while each sampling cell contains $2\times2\times1$ collision cells,
or $N_{p}=101$ particles on average. The height of the box is $L_{y}=256\lambda=960$
and the imposed concentrations at the walls are $c_{B}=0.25$ and
$c_{T}=0.75$. For System B, the DSMC parameters and $c_{T/B}$ are
the same as for to System A, but the sampling cells are twice as large,
$4\times4\times1$ collision cells each, and the system height is
twice as large, $L_{y}=512\lambda=1920$. We obtain similar results
using twice smaller collision cells (not shown). For the quasi two-dimensional
systems, the thickness is $L_{z}=7.5=2\lambda$ and there is only
one DSMC collision cell along the $z$ direction. Figure \ref{fig:j1_vs_Nx}a
shows that $\chi_{\text{eff}}$ grows like $\ln L_{x}$, with a slope
that is well-predicted by Eq. (\ref{eq:chi_eff_2D}). For widths larger
than about $8$ mean free paths, $\chi_{0}$ becomes constant and
rather similar to the kinetic theory prediction. It is important to
point out that $\chi_{0}$ is not a fundamental material constant
and in fact depends on the shape of the sampling cells (see Section
\ref{sub:chi_0}).

In Fig. \ref{fig:j1_vs_Nx}b we show results from three dimensional
DSMC simulations, in which the system width ($x$) and depth ($z$)
directions are equivalent, $L_{z}=L_{x}=L$, and the rest of the parameters
are the same as for System A. Similar behavior is seen as in two dimensions,
except that now the effective diffusion grows as $-L^{-1}$ and saturates
to a constant value for large $L$, assuming that $L_{y}\gg L$.

\subsubsection{\label{sub:FiniteHeight}Corrections due to finite height}

The predictions of the simplified fluctuating hydrodynamic theory,
Eqs. (\ref{eq:chi_eff_2D}) and (\ref{eq:chi_eff_3D}), are shown
in Fig. \ref{fig:j1_vs_Nx} and seen to be in very good agreement
with the particle simulations for intermediate $L_{x}$. However,
the particle data shown in Fig. \ref{fig:j1_vs_Nx}a shows measurable
deviations from the simple theory for $L_{x}\gtrsim L_{y}/2$. To
understand the discrepancy, recall that the incompressible isothermal
theory assumed that $L_{y}$ is essentially infinite and thus in two
dimensions $\chi_{\text{eff}}$ grows unbounded in the macroscopic
limit. A scaling analysis suggests a modification of (\ref{eq:chi_eff_2D})
to account for the finite height of the system,\begin{equation}
\chi_{\text{eff}}^{(2D)}\approx\chi+\frac{k_{B}T}{4\pi\rho(\chi+\nu)L_{z}}\,\left[\ln\frac{L_{x}}{L_{0}}-f(r)\right],\label{eq:chi_eff_2D_finite}\end{equation}
where $r=L_{x}/L_{y}$ is the aspect ratio of the system, and $f(r)$
is some function that is close to zero for small $r$ and grows asymptotically
as $\ln(r)$. Therefore, when $L_{x}\gg L_{y}$, $\chi_{\text{eff}}$
saturates to a constant value that grows as $\ln(L_{y}/L_{0})$.

One can extend the theoretical calculations to account for the hard
wall boundary conditions in the $y$ direction \citet{FluctHydroNonEq_Book},
however, such a calculation is non trivial. Instead, we have used
the finite-volume solver developed in Ref. \citet{LLNS_S_k} to solve
the non-linear system of SPDEs (\ref{LLNS_primitive}) for the same
system dimensions as in the particle simulations. The results, shown
in Fig. \ref{fig:j1_vs_Nx}, are in excellent agreement with the particle
simulations for the larger system sizes. Note that while our SPDE
solver includes all of the nonlinear terms in (\ref{LLNS_primitive}),
we may artificially reduce the amplitude of the noise and thus the
magnitude of the fluctuations by some factor $\epsilon\ll1$. This
reduces the effect of the nonlinearities and effectively gives a quasi-linearized
finite-volume SPDE solver. The advective mass flux due to the velocity
fluctuations can be estimated as \begin{equation}
\D{\bar{j}_{\parallel}^{(1)}}(y)\approx\epsilon^{-2}\av{\left(\d{\rho_{1}}\right)\left(\d v_{\parallel}^{(1)}\right)}\approx\epsilon^{-2}\bar{\rho}\av{\left(\d c\right)\left(\d v_{\parallel}\right)},\label{eq:dj1_y_def}\end{equation}
and may depend on $y$ especially close to the walls or when $L_{y}\gtrsim L_{x}$.
The sum of the average diffusive and advective mass fluxes must be
independent of $y$,\begin{equation}
\bar{j}_{\parallel}^{(1)}=\bar{\rho}\chi_{\text{eff}}\,\frac{c_{T}-c_{B}}{L_{y}}=\bar{\rho}\chi_{0}\frac{d\bar{c}(y)}{dy}-\D{\bar{j}_{\parallel}^{(1)}}(y),\label{eq:j1_bar_const}\end{equation}
which implies that the macroscopic concentration profile $\bar{c}(y)$
is affected by the fluctuations as well and cannot be strictly linear.
From the conditions $c(0)=c_{B}$ and $c(L_{y})=c_{T}$ and (\ref{eq:j1_bar_const})
we obtain the relation\[
\chi_{\text{eff}}=\chi_{0}-\left[\bar{\rho}\left(c_{T}-c_{B}\right)\right]^{-1}\;\int_{y=0}^{L_{y}}\left[\D{\bar{j}_{\parallel}^{(1)}}(y)\right]dy,\]
which is how we calculate the effective diffusion coefficient from
the numerical SPDE solution. We have verified that the results are
independent of $\epsilon$ to within statistical accuracy for $\epsilon\leq1/2$.

\begin{figure}[h]
\begin{centering}
\includegraphics[width=0.75\columnwidth]{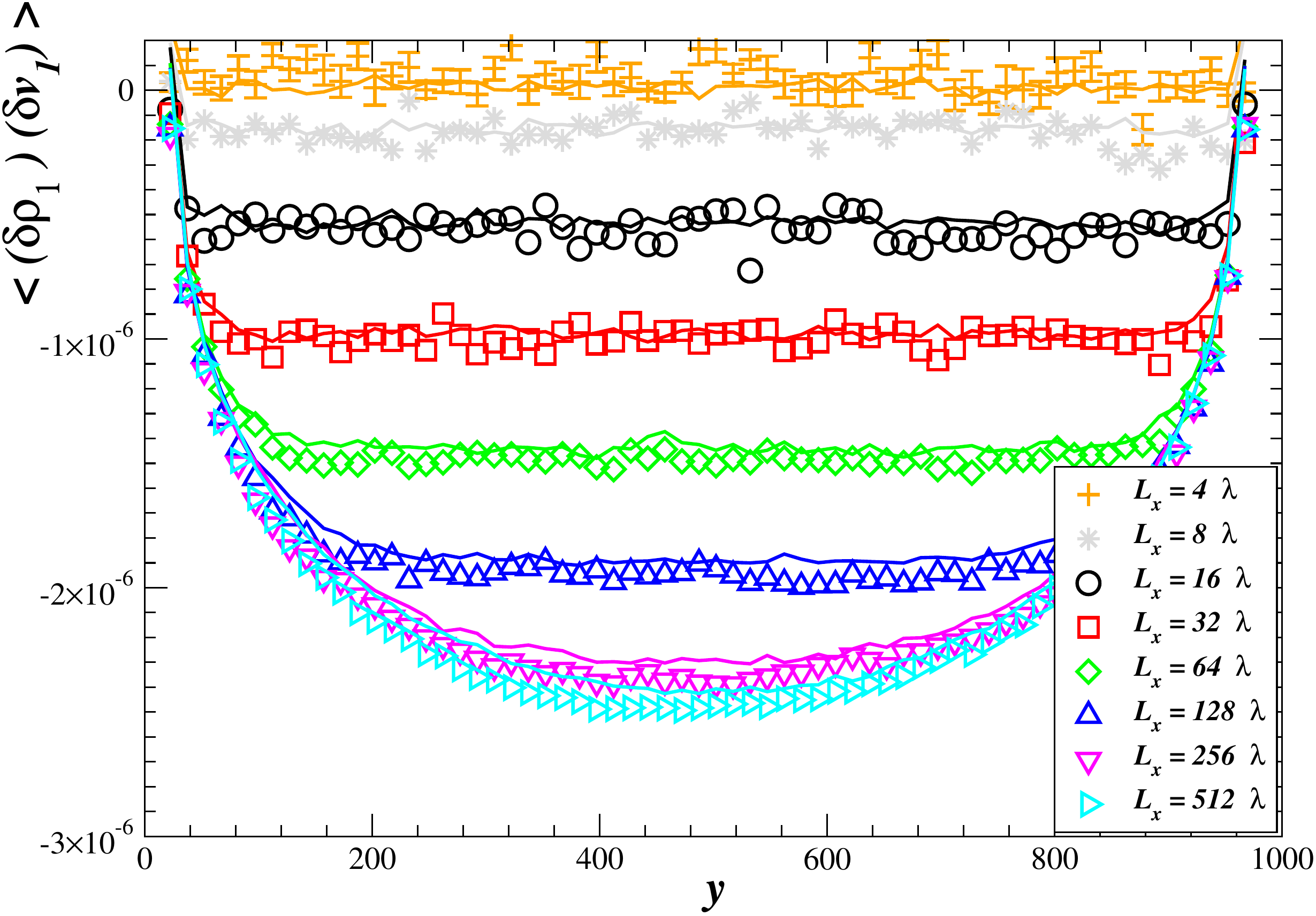}
\par\end{centering}

\caption{\label{fig:LLNS-2}The fluctuating contribution to the mean diffusive
flux, $\av{\left(\d{\rho_{1}}\right)\left(\d v_{\parallel}^{(1)}\right)}$,
as a function of the height of the sampling cell $y$, for several
system widths $L_{x}$, keeping $L_{y}=256\lambda$. Data from quasi-two
dimensional particle simulations (System A) is shown with symbols,
and lines of the same color show data for $\bar{\rho}\av{\left(\d c\right)\left(\d v_{\parallel}\right)}$
obtained using a two-dimensional finite-volume solver \citet{LLNS_S_k}
for the LLNS equations (\ref{LLNS_primitive}). The error bars for
the particle data are shown for $L_{x}=4\lambda$, and are similar
or smaller for the remaining systems and for the finite-volume results.}

\end{figure}

The velocity-concentration correlation $\D{\bar{j}_{\parallel}^{(1)}}(y)$
obtained from the finite-volume solver is shown in Fig. \ref{fig:LLNS-2},
along with the corresponding particle data for comparison. Excellent
agreement is seen, demonstrating that the finite-volume solution correctly
accounts for the influence of the boundaries. Note that the partial
velocity $\V v_{1}$ is not included as an independent variable in
(\ref{LLNS_primitive}) and the mean velocity $\V v$ is used instead.
When compressibility effects are included, $\V v$, unlike $\V v_{1}$,
is correlated with $\rho_{1}$ even in one dimension. This makes a
direct comparison between the effective diffusion coefficient in the
particle and finite-volume simulations difficult. However, the dependence
of $\chi_{\text{eff}}$ on system size should be the same in both
types of simulations, once the bare transport coefficient is adjusted
empirically.

\subsubsection{\label{sub:chi_0}Renormalized Diffusion Coefficient}

The renormalized diffusion coefficient $\chi_{0}$ in the Fickian
diffusive flux is an input to the SPDE calculations and assumed constant.
In our calculations we used the prediction of kinetic theory \citet{DSMC_CellSizeError,DSMC_TimeStepError},
also shown in Fig. \ref{fig:j1_vs_Nx}. In finite-volume solvers,
the spacing of the computational grid plays the equivalent of the
cutoff length $L_{\text{mol}}$, and therefore the effective mass
flux depends on the grid spacing. Furthermore, there are numerical
grid artifacts in the SPDE solution at length and time scales comparable
to the numerical discretization parameters \citet{LLNS_S_k}. To correct
for these errors, we have added a constant to the effective diffusion
coefficient obtained from SPDE runs to match $\chi_{\text{eff}}$
from the particle simulations for $L_{x}=L_{0}=16\lambda$. This correction
essentially renormalizes $\chi_{0}$ based on the size of the finite-volume
hydrodynamic cells.

One can think of $\chi_{0}$ defined via (\ref{eq:bare_definition})
as the fluctuation-renormalized diffusion coefficient at length scales
determined by the shape of the sampling or observation volume $\D{\mathcal{V}}$.
In this sense, $\chi_{0}$ is the physical-space equivalent of the
wavenumber-dependent diffusion coefficient $\chi\left(\V k,\omega=0\right)$
commonly used in linear response theories \citet{DiffusionRenormalization_I,SelfDiffusion_Linearity}.
A theoretical prediction for $\chi_{0}$ can be obtained by starting
from linearized theory for the fluctuating fields $\rho_{1}\left(\V r,t\right)=\bar{\rho}_{1}+\d{\rho_{1}}\left(\V r,t\right)$
and $\V v_{1}\left(\V r,t\right)=\bar{\V v}_{1}+\d{\V v_{1}}\left(\V r,t\right)$.
The instantaneous velocity in a given sampling cell was defined through
the instantaneous momentum density $\V j_{1}=\rho_{1}\V v_{1}$ averaged
over the sampling cell,\[
\av{\rho_{\D{\mathcal{V}}}^{(1)}}\av{\V v_{\D{\mathcal{V}}}^{(1)}}=\bar{\rho}_{1}\av{\V v_{\D{\mathcal{V}}}^{(1)}}=\bar{\rho}_{1}\left\langle \frac{\int_{\D{\mathcal{V}}}\rho_{1}\V v_{1}\, d\V r}{\int_{\D{\mathcal{V}}}\rho_{1}\, d\V r}\right\rangle =\av{\rho_{\D{\mathcal{V}}}^{(1)}\V v_{\D{\mathcal{V}}}^{(1)}}-\D{\V j}_{F},\]
where to second order in the fluctuations,\begin{align*}
\D{\V j}_{F} & =\D V^{-2}\int_{\D{\mathcal{V}}}d\V r\,\int_{\D{\mathcal{V}}}d\V r^{\prime}\,\left\langle \rho_{1}(\V r,t)\V v_{1}(\V r^{\prime},t)\right\rangle \\
= & \D V^{-2}\int_{\D{\mathcal{V}}}d\V r\,\int_{\D{\mathcal{V}}}d\V r^{\prime}\;\left(2\pi\right)^{-6}\int_{\V k}d\V k\int_{\V k^{\prime}}d\V k^{\prime}\;\av{\widehat{\delta\rho_{1}}\left(\V k,t\right)\widehat{\delta\V v_{1}}^{\star}\left(\V k^{\prime},t\right)}e^{i\left(\V k\cdot\V r-\V k^{\prime}\cdot\V r^{\prime}\right)}\\
= & \left(2\pi\right)^{-3}\int_{\V k}\left[\D V^{-2}\int_{\D{\mathcal{V}}}d\V r\,\int_{\D{\mathcal{V}}}d\V r^{\prime}\, e^{i\V k\cdot\left(\V r-\V r^{\prime}\right)}\right]\mathcal{S}_{\rho_{1},\V v_{1}}\left(\V k\right)\, d\V k\\
= & \left(2\pi\right)^{-3}\int_{\V k}F_{\D{\mathcal{V}}}\left(\V k\right)\,\mathcal{S}_{\rho_{1},\V v_{1}}\left(\V k\right)\, d\V k,\end{align*}
and $F_{\D{\mathcal{V}}}\left(\V k\right)$ is the low pass filter
that already appeared in Eq. (\ref{eq:F_x_filt}). The result of this
calculation {[}c.f. Eq. (\ref{eq:dchi_k_int}){]},\begin{align}
\chi_{\text{eff}} & =\chi-\left(2\pi\right)^{-3}\int_{\V k}\left[\rho_{0}^{-1}\D{\mathcal{S}}_{\rho_{1},v_{\Vert}^{(1)}}\left(\V k\right)\right]\, d\V k\nonumber \\
\chi_{0} & =\chi-\left(2\pi\right)^{-3}\int_{\V k}\left[1-F_{\D{\mathcal{V}}}\left(\V k\right)\right]\left[\rho_{0}^{-1}\D{\mathcal{S}}_{\rho_{1},v_{\Vert}^{(1)}}\left(\V k\right)\right]\, d\V k,\label{eq:chi_definitions}\end{align}
shows that $\chi_{\text{eff}}$ includes contributions from all wavenumbers
present in the system, while $\chi_{0}$ only includes {}``sub-grid''
contributions, from wavenumbers larger than $2\pi/\D x$. The theoretical
predictions shown in Fig. \ref{fig:j1_vs_Nx} are based on numerically
evaluating (\ref{eq:chi_definitions}) after replacing the integrals
over $k_{x}$ and $k_{z}$ (\ref{eq:chi_definitions}) with the appropriate
sums, assuming $\mathcal{S}_{\rho_{1},\V v_{1}}\approx\bar{\rho}\mathcal{S}_{c,\V v}$
and using Eq. (\ref{eq:S_c_vy}). The bare diffusion coefficient $\chi$
is adjusted so that $\chi_{\text{eff}}$ matches the particle result
for $L_{x}=L_{0}=16\lambda$, and good agreement is observed between
(\ref{eq:chi_definitions}) and the particle data for $\chi_{0}$
for all but the smallest $L_{x}$.

While it is intuitive to expect that the bare diffusion coefficient
should account for molecular, or non-hydrodynamic, degrees of freedom,
the division $\chi_{\text{eff}}=\chi+\D{\chi}$ is arbitrary, and
in fact there is no unambiguous way to define $\chi$. This is evident
in the theory because of the need to introduce an ad-hoc molecular
cutoff as a way to separate the {}``microscopic'' from the {}``mesoscopic''
scales. By contrast, the \emph{locally renormalized} diffusion coefficient
$\chi_{0}$ defined in (\ref{eq:bare_definition}) explicitly depends
on the size of the sampling (hydrodynamic) cells $\D V=\abs{\D{\mathcal{V}}}$
used to define the hydrodynamic quantities from the particle configuration.
Combining the two equations in (\ref{eq:chi_definitions}) gives the
renormalization relation at large scales, \[
\chi_{\text{eff}}=\chi_{0}\left(\D{\mathcal{V}}\right)-\left(2\pi\right)^{-3}\int_{\V k}F_{\D{\mathcal{V}}}\left(\V k\right)\left[\rho_{0}^{-1}\D{\mathcal{S}}_{\rho_{1},v_{\Vert}^{(1)}}\left(\V k\right)\right]d\V k,\]
which eliminates the dependence on the ad-hoc cutoff wavenumber since
$F_{\D{\mathcal{V}}}$ filters contributions from large wavenumbers,
at least within the simple perturbative (quasi-linear) theory.

\section{\label{sec:VACF}Connections to Earlier Work}

While our computer simulations are the first hydrodynamic study of
the dependence of transport on system size, there is a substantial
body of literature that has discussed the effect from a theoretical
perspective or studied smaller particle systems. In this section we
explicitly connect our analysis to previous approaches, and discover
direct relations with work that might have, at first sight, been assumed
to be unrelated.

\subsection{\label{sub:VACF_MD}Relation to Long-Time Tails}

It is well known that the self-diffusion coefficient is given by the
integral of the equilibrium velocity autocorrelation function (VACF)
$C(t)$ of the fluid particles \citet{LongRangeCorrelations_MD}.
The long-time tail of $C(t)$ has been extensively studied in the
literature both computationally and through several theories, including
heuristic hydrodynamic arguments \citet{Alder:70,VACFTail_Widom},
kinetic theory \citet{VACFTail_Cohen} and (second-order) mode-mode
coupling hydrodynamic theory \citet{ModeModeCoupling}. Ultimately
all derivations give the same result including not just the power-law
dependence but also the coefficient of the tail, specifically, in
three dimensions $C(t)\approx k_{B}T/\left\{ 12\rho\left[\pi\left(\chi+\nu\right)t\right]^{3/2}\right\} $,
while for quasi-two dimensional systems $C(t)\approx k_{B}T/\left[8\pi\rho L_{z}\left(\chi+\nu\right)t\right]$.

A crucial point is that the VACF explicitly depends on the system
size due to periodic boundaries, and so its integral, which gives
the diffusion coefficient, also depends on system size. More explicitly,
ignoring acoustic effects, the VACF has the power-law dependence only
for $L_{\text{mol}}^{2}/\left(\chi+\nu\right)\ll t\ll L^{2}/\left(\chi+\nu\right)$,
and it decays exponentially for large times \citet{BrownianParticle_SIBM}.
Ignoring prefactors, the contribution of the tail to the diffusion
coefficient in three dimensions is estimated as\begin{equation}
\D{\chi}_{\text{tail}}\sim\int_{L_{\text{mol}}^{2}/\left(\chi+\nu\right)}^{L^{2}/\left(\chi+\nu\right)}\frac{k_{B}T}{\rho\left[\left(\chi+\nu\right)t\right]^{3/2}}dt\sim\frac{k_{B}T}{\rho\left(\chi+\nu\right)}\left(\frac{1}{L_{\text{mol}}}-\frac{1}{L}\right),\label{eq:dchi_VACF_est}\end{equation}
which of exactly the same form as (\ref{eq:chi_eff_3D}). A similar
calculation in two dimensions reproduces the logarithmic dependence
in (\ref{eq:chi_eff_2D}).

A more quantitative comparison to the theories for the VACF tail can
be made by examining the predictions of the mode-mode coupling theory
for the long-time tail, reviewed in detail in Section 3.2 of Ref.
\citet{ModeModeCoupling}. The relevant formula for the VACF is their
Eq. (3.39), which, after integrating over the Boltzmann velocity distribution,
becomes\[
C(t)=\frac{k_{B}T}{12\pi^{3}\rho}\int_{\V k}\exp\left[-\left(\chi+\nu\right)k^{2}t\right]d\V k.\]
In \citet{ModeModeCoupling}, the integral over $\V k$ is performed
assuming an infinite system and the time dependence kept in order
to see the behavior of the tail at long times. If we integrate over
$t$ instead, we get\begin{equation}
\D{\chi}_{\text{tail}}=\int_{t=0}^{\infty}C(t)dt=\frac{k_{B}T}{12\pi^{3}\rho(\chi+\nu)}\int_{\V k}k^{-2}d\V k,\label{eq:dchi_VACF}\end{equation}
which is seen to identical to the integral in Eq. (\ref{eq:dchi_k_int})
under the assumption that all three directions $x$, $y$ and $z$
are identical (as done in all VACF calculations), \[
\D{\chi}=\frac{k_{B}T}{(2\pi)^{3}\rho\left(\chi+\nu\right)}\;\int_{\V k}\frac{k_{x}^{2}+k_{z}^{2}}{k^{4}}\, d\V k=\frac{k_{B}T}{(2\pi)^{3}\rho\left(\chi+\nu\right)}\;\frac{2}{3}\int_{\V k}k^{-2}\, d\V k=\D{\chi}_{\text{tail}}.\]
Note that for a finite system one ought to replace the integrals over
$\V k=\V{\kappa}\cdot2\pi/L$ with sums over $\V{\kappa}\in Z^{d}$
that exclude $\V{\kappa}=0$,\begin{equation}
\D{\chi}_{\text{tail}}=\frac{2k_{B}T}{3\rho(\chi+\nu)L^{3}}\sum_{\V{\kappa}\neq\V 0}k^{-2}.\label{eq:dchi_VACF_sum}\end{equation}

In the Molecular Dynamics (MD) literature, the dependence on $L^{-1}$
in Eq. (\ref{eq:dchi_VACF_est}) is considered a finite-size effect
that ought to be removed in order to extract the bulk ($L\rightarrow\infty$)
limit of the diffusion coefficient \citet{TracerDiffusion_HS,SelfDiffusionFinite_HS,FiniteSize_Diffusion_MD}.
A hydrodynamic theory for the finite-size correction, based on the
Oseen tensor for a finite periodic system, has been developed several
times \citet{PolymerChainMD_Dunweg} and is confirmed numerically
in Refs. \citet{FiniteSize_Diffusion_MD}. This theory focuses on
viscous effects only, and we will thus replace $\nu$ with $\nu+D$
in Eqs. (10,11) in Ref. \citet{FiniteSize_Diffusion_MD}, to obtain\[
\D{\chi}_{\text{MD}}=\frac{k_{B}T}{6\pi\left(\chi+\nu\right)}\left[\tilde{f}(k_{\max})+\sum_{\V{\kappa}\neq\V 0}4\pi L^{-3}k^{-2}\exp\left(-\frac{k^{2}}{4k_{\max}^{2}}\right)\right],\]
where $\tilde{f}(k)$ is some function. Assuming that $k_{\max}$
is large, the system-size dependence is captured in the last term,
which is exactly the same as Eq. (\ref{eq:dchi_VACF_sum}),\[
\D{\chi}_{\text{MD}}\approx\frac{k_{B}T}{6\pi\left(\chi+\nu\right)}\sum_{\V{\kappa}\neq\V 0}4\pi L^{-3}k^{-2}=\D{\chi}_{\text{tail}}=\D{\chi}.\]

\begin{figure}[h]
\begin{centering}
\includegraphics[width=0.49\columnwidth]{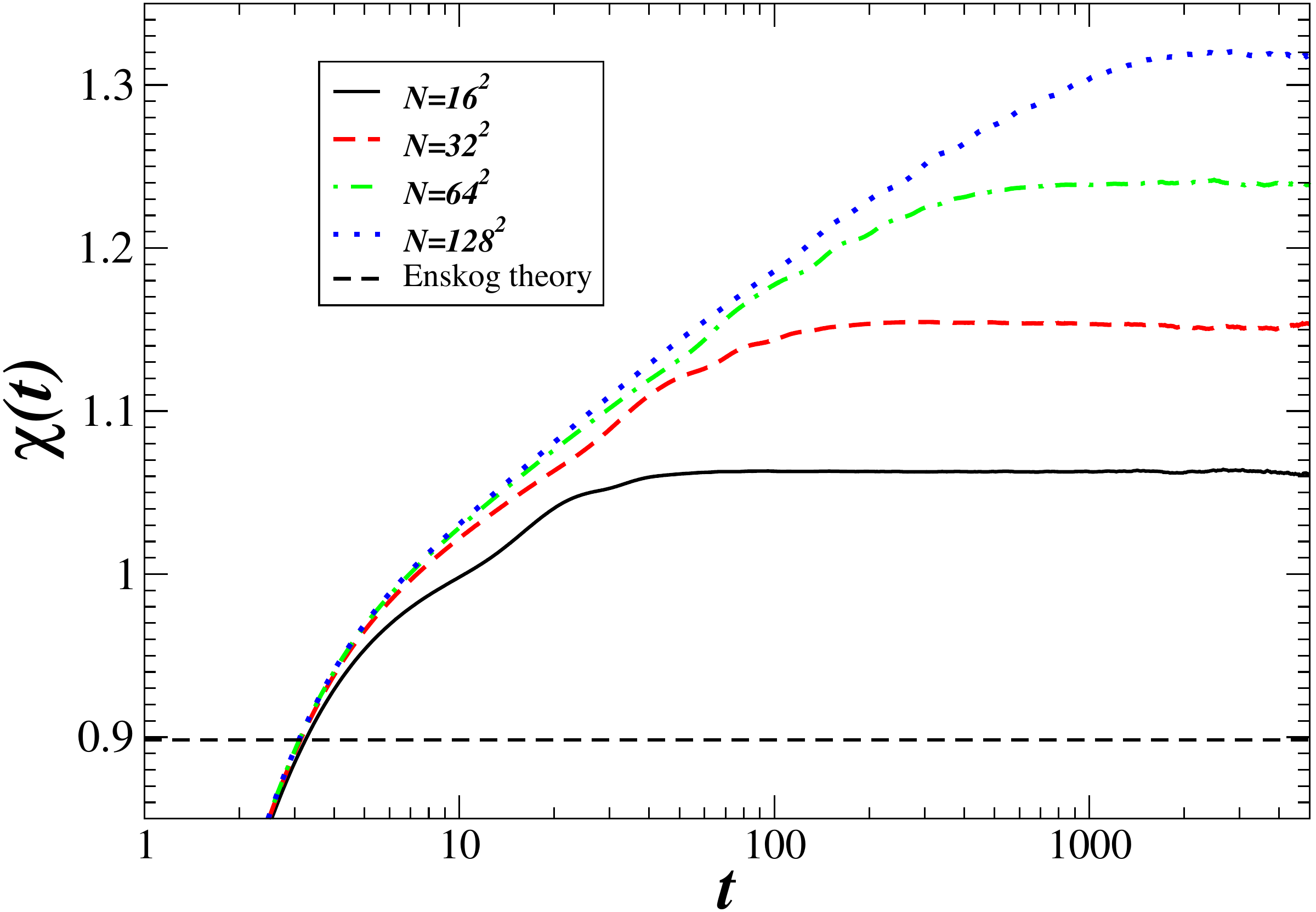}\includegraphics[width=0.49\columnwidth]{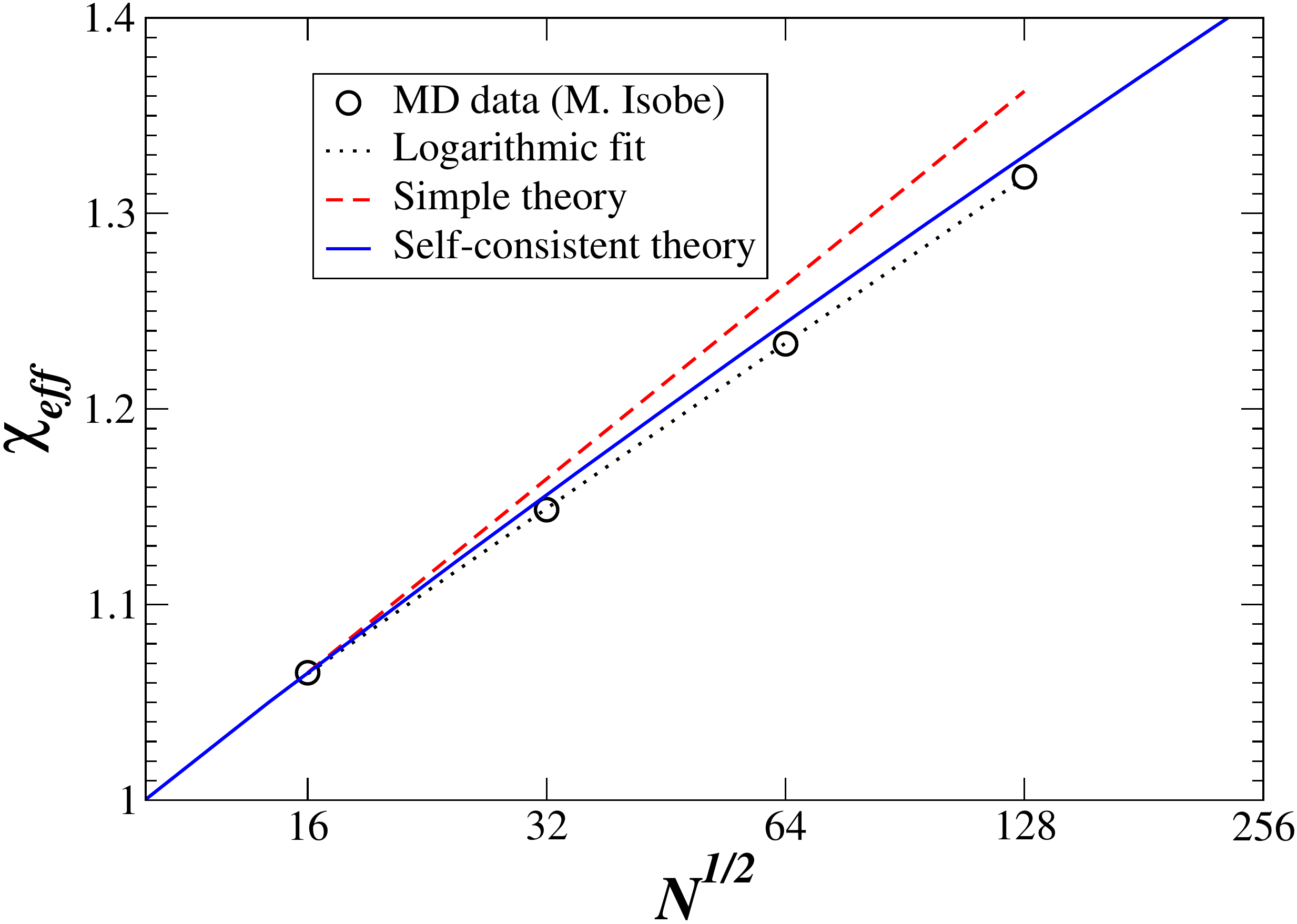}
\par\end{centering}

\caption{\label{fig:D_HS2D}System-size dependence of the diffusion coefficient
for hard disks at volume fraction $\phi=0.18$ (data courtesy of Masaharu
Isobe \citet{VACF_2D_HS}). Time and lengths are measured in natural
units, or, equivalently, $k_{B}T=1$, $m=1$ and disk diameter $\sigma=1$.
(Left) The time-dependent diffusion coefficient $\chi(t)$, as determined
from a numerical integral of the velocity autocorrelation function,
for several system sizes (see legend). (Right) The observed increase
of the effective diffusion coefficient with system size (symbols),
as obtained from the large-time limit of $\chi(t)$. Theoretical predictions
are shown for comparison, for both the simple theory (\ref{eq:chi_eff_MD})
(dashed line), as well as for the self-consistent theory (\ref{eq:chi_eff_2D_SC})
with the system size $N_{0}=16^{2}$ used to define a reference length
$L_{0}$. A logarithmic fit with an adjustible slope is also shown
(dotted line).}

\end{figure}

There are few molecular dynamics studies of the system size dependence
of the diffusion coefficient in sufficiently large two-dimensional
systems. Isobe has performed one of the most extensive hard-disk molecular
dynamics studies of the hydrodynamic tail of the VACF \citet{VACF_2D_HS}.
We have performed numerical integration of $C(t)$ using the data
of Isobe for hard disks at packing fraction $\phi=0.18$, for square
systems sizes containing from $N=16^{2}$ to $N=128^{2}$ hard disks.
For these parameters the statistical accuracy of the data appears
sufficient to resolve the asymptotic plateau $\chi_{\text{eff}}=\lim_{\tau\rightarrow\infty}\chi(t)$
of the time-dependent diffusion coefficient\[
\chi(t)=\int_{\tau=0}^{t}C(\tau)d\tau,\]
as illustrated in Fig. \ref{fig:D_HS2D}. Since the aspect ratio of
all of Isobe's simulations is fixed at $L_{x}/L_{y}=1$, Eq. (\ref{eq:chi_eff_2D_finite})
suggests that the difference between $\chi_{\text{eff}}$ for a system
size of $N$ disks and the smallest system of $N_{0}=16$ disks is\begin{equation}
\chi_{\text{eff}}(N)-\chi_{\text{eff}}(N_{0})\approx\frac{k_{B}T}{8\pi\rho(\chi+\nu)}\,\ln\frac{N}{N_{0}}.\label{eq:chi_eff_MD}\end{equation}
In Fig. \ref{fig:D_HS2D} we compare this prediction to the numerical
integral of Isobe's data, using Enskog kinetic theory \citet{Enskog_2DHS}
values for the {}``bare'' transport coefficients $\chi$ and $\nu$
of the hard disk fluid at this density. Good agreement is observed,
demonstrating that the effect we observe is not an artifact of DSMC
but rather a generic property of fluids.

\subsection{Self-Consistent Theory}

The theoretical predictions with which we compared the particle results
were based on a leading-order perturbative theory \citet{DiffusionRenormalization_I}
that relies on the solution of the linearized equations of fluctuating
hydrodynamics. A fully nonlinear theory, however, remains illusive.
In the context of infinite (bulk) systems, a systematic perturbative
theory that accounts for corrections of order higher than quadratic
in the fluctuations has been discussed in Refs. \citet{DiffusionRenormalization_II,TrilinearModeCoupling}.
In three dimensions, the conclusion of such studies has been that
the higher-order terms do not affect the form of (\ref{eq:chi_definitions}).
In two dimensions, several calculations \citet{VACF_2D_SelfConsistentMC,TrilinearModeCoupling,SCModeModeCoupling2D}
and numerical simulations \citet{VACF_2Divergence,VACF_2D_HS} suggest
that including higher-order terms changes the logarithmic growth in
(\ref{eq:chi_eff_2D}). Specifically, it has been predicted that the
self-consistent power-law decay for the VACF is faster than $t^{-1}$,
$C(t)\sim\left(t\sqrt{\ln t}\right)^{-1}$, which changes the asymptotic
growth of $\chi_{\text{eff}}$ from $\ln L_{x}$ to $\sqrt{\ln L_{x}}$.

In order to obtain a self-consistent form of (\ref{eq:chi_eff_2D}),
we reconsider the derivation in Section \ref{sub:Two-Dimensions}.
The cause of the diffusion growth from system width $L_{x}$ to $L_{x}+dL_{x}$
is the added contribution to the integral in Eq. (\ref{eq:j_lnLx})
from wavenumbers in the band $\abs{k_{x}}\in2\pi L_{x}^{-1}\cdot\left[1-dL_{x}/L_{x},1\right]$.
If we postulate that the concentration fluctuations at wavenumber
$k_{x}=2\pi/L_{x}$ evolve with the renormalized diffusion coefficient
$\chi_{\text{eff}}\left(L_{x}\right)$, instead of the bare one, we
obtain an ordinary differential equation for $\chi_{\text{eff}}\left(L_{x}\right)$,
\begin{equation}
\frac{d\chi_{\text{eff}}}{dL_{x}}\approx\frac{k_{B}T}{(2\pi)^{2}\rho\left(\chi_{\text{eff}}+\nu\right)L_{z}}\left(2\,\frac{k_{x}}{L_{x}}\right)\int_{k_{y}}\frac{k_{x}^{2}}{\left(k_{x}^{2}+k_{y}^{2}\right)^{2}}\, dk_{y}=\frac{k_{B}T}{4\pi\rho(\chi_{\text{eff}}+\nu)L_{z}}L_{x}^{-1},\label{eq:dchi_dLx_2D}\end{equation}
where we have assumed that viscosity does not change substantially
with system size (consistent with existing molecular dynamics data).
Solving the differential equation (\ref{eq:dchi_dLx_2D}) with the
condition $\chi_{\text{eff}}\left(L_{0}\right)=\chi$ leads to a diffusion
enhancement that grows slightly slower than logarithmically,\begin{equation}
\chi_{\text{eff}}\left(L_{x}\right)\approx\left(\chi+\nu\right)\left[1+\frac{k_{B}T}{2\pi\rho(\chi+\nu)^{2}L_{z}}\,\ln\frac{L_{x}}{L_{0}}\right]^{1/2}-\nu.\label{eq:chi_eff_2D_SC}\end{equation}
Although the self-consistent form (\ref{eq:chi_eff_2D_SC}) still
diverges with system length, it is important to observe that for any
finite system $\chi_{\text{eff}}$ is well-defined. The predictions
of (\ref{eq:chi_eff_2D_SC}) for a hard-disk system at packing fraction
$\phi=0.18$ are shown in Fig. \ref{fig:D_HS2D}. We have used the
Enskog kinetic theory for the viscosity, which is in good agreement
with published molecular dynamics data at this density, and set $\rho L_{z}$
to be the mass density in the plane. The self-consistent theory is
seen to be in better agreement with the molecular dynamics data than
the simple theory (\ref{eq:chi_eff_2D}); however, the difference
between the two is small for the system sizes at which we presently
have reliable data for the effective diffusion coefficient.

Repeating the same self-consistent calculation in three dimensions
gives the self-consistent form,\begin{equation}
\chi_{\text{eff}}\left(L\right)\approx\left(\chi+\nu\right)\left[1+\frac{\alpha\, k_{B}T}{\rho(\chi+\nu)^{2}}\left(\frac{1}{L_{0}}-\frac{1}{L}\right)\right]^{1/2}-\nu,\label{eq:chi_eff_3D_SC}\end{equation}
which happens to be the solution of the consistency condition\[
\chi_{\text{eff}}=\chi+\frac{\alpha\, k_{B}T}{\rho\left[\nu+(\chi+\chi_{\text{eff}})/2\right]}\left(\frac{1}{L_{0}}-\frac{1}{L}\right),\]
reminiscent of the form obtained in Ref. \citet{DiffusionRenormalization_II}
except that the arithmetic average of $\chi$ and $\chi_{\text{eff}}$
appears in the denominator instead of just $\chi_{\text{eff}}$. In
three dimensions, the difference between the self-consistent (\ref{eq:chi_eff_3D_SC})
and the simple (\ref{eq:chi_eff_3D}) theories is very small and thus
rather difficult to observe computationally.

In both two and three dimensions, the self-consistent predictions
(\ref{eq:chi_eff_2D_SC},\ref{eq:chi_eff_3D_SC}) will only deviate
from the simple theory (\ref{eq:chi_eff_2D},\ref{eq:chi_eff_3D})
substantially when the diffusion enhancement becomes comparable to
the bare coefficient, that is, when hydrodynamic effects become comparable
to molecular ones. The estimates presented in Appendix \ref{AppendixEstimates}
show that reaching the regime where $\D{\chi}\gtrsim\chi$ is difficult
to achieve with particle simulations. In nonlinear fluctuating hydrodynamics
finite-volume solvers \citet{LLNS_S_k}, one has the freedom to choose
the various parameters so as to make the effect of advection by velocity
fluctuations much more prominent, similarly to what is done in Ref.
\citet{VACF_2Divergence} using a two-dimensional lattice gas method.
Experimentally, two dimensional systems can be realized by using thin
films, for example, liquid or liquid crystal films. In liquid films,
however, velocity fluctuations below a certain cutoff wavenumber are
suppressed because of the drag by the surrounding fluid, and therefore
the diffusion enhancement saturates for systems much larger than the
cutoff length scale \citet{GiantFluctuations_ThinFilms}.

\section{Conclusions and Future Directions}

The results of our particle simulations confirm that fluctuating hydrodynamics
is a powerful tool for understanding transport at small scales. Our
results conclusively demonstrate that the advection by thermal velocity
fluctuations affects the \emph{mean} transport in nonequilibrium finite
systems and thus the advective nonlinearities, such as the term $\left(\delta c\right)\left(\d{\V v}\right)$
in (\ref{eq:dc_t_nonlinear}), ought to be retained in the equations
of fluctuating hydrodynamics. We demonstrated explicitly that the
correction to the bare or molecular transport coefficients due to
the VACF tail \citet{LongRangeCorrelations_MD}, hydrodynamic interactions
with periodic images of a given particle \citet{FiniteSize_Diffusion_MD},
and the contribution due to thermal fluctuations \citet{DiffusionRenormalization_I,ExtraDiffusion_Vailati}
studied here, are all the same physical phenomenon simply calculated
through different theoretical approaches, all of which are equivalent
because of linearity. The advantage of fluctuating hydrodynamics is
that it is simple, and it can take into account the proper boundary
conditions and exact geometry, especially if a numerical SPDE solver
is used. Furthermore, other effects such as gravity \citet{ExtraDiffusion_Vailati},
temperature variations \citet{TemperatureGradient_Cannell}, or time
dependence \citet{GiantFluctuations_Theory,GiantFluctuations_ThinFilms},
can easily be included. It remains as a future challenge to verify
the predictions of fluctuating hydrodynamics for the effect of fluctuations
on diffusive transport in spatially non-uniform systems, either through
particle simulations or laboratory experiments \citet{GiantFluctuations_ThinFilms}.

Renormalization has often been invoked as a way to fold the contribution
from fluctuations into the effective transport coefficients, however,
this only works in three dimensions for very large systems. In two
dimensions, a macroscopic limit does not exist, and in three dimensions
there are strong finite-size corrections even for systems with dimensions
much larger than molecular scales. Theoretical modeling of finite
systems at the nano or microscale thus requires including nonlinear
hydrodynamic fluctuations. However, a complete nonlinear theory has
yet to be developed, and requires detailed understanding of the role
of large wavenumber cutoffs (regularizations) that are necessary to
make the SPDEs well-behaved. Furthermore, the proper physical and
mathematical interpretation of other types of nonlinearities in (\ref{LLNS_primitive})
and (\ref{eq:LLNS_incomp_v},\ref{eq:LLNS_incomp_c}), notably the
dependence of the transport coefficients and the stochastic forcing
amplitude on the fluctuations, remain to be clarified. Future work
should also study momentum and heat transfer in steady states, as
well as time-dependent transport in systems that are far from equilibrium.
\begin{acknowledgments}
We are grateful to Masaharu Isobe for sharing his hard-disk MD data
and helping us analyze it. We thank Berni Alder, Doriano Brogioli,
Jonathan Goodman and Eric Vanden-Eijnden for informative discussions
and helpful suggestions on improving this work. This work was supported
in part by the DOE Applied Mathematics Program of the DOE Office of
Advanced Scientific Computing Research under the U.S. Department of
Energy under contract No. DE-AC02-05CH11231.
\end{acknowledgments}
\begin{appendix}

\section{\label{AppendixEstimates}Estimates of the Diffusion Enhancement}

It is instructive to do some scaling analysis of the order of magnitude
of $\D{\chi}$ in realistic fluid systems. Following (\ref{eq:chi_eff_3D}),
the hydrodynamic contribution to the diffusion coefficient for a large
three dimensional system is estimated as\begin{equation}
\D{\chi}\sim\frac{k_{B}T}{\rho(\chi+\nu)L_{\text{mol}}},\label{eq:dchi_est_3D}\end{equation}
For gases, $\D{\chi}$ can be estimated by using Chapman-Enskog values
for the transport coefficients for a hard-sphere gas with molecular
collision diameter $\sigma\approx L_{\text{mol}}$, specifically,
$\chi\sim\nu\sim\left(\rho\sigma^{2}\right)^{-1}\sqrt{mk_{B}T}$.
For liquids, the Schmidt number is large, $S_{c}=\nu/\chi>10^{2}$,
and Stokes-Einstein's relation suggests that $\chi\sim k_{B}T/\left(\rho\nu\sigma\right)$.
For both gases and liquids we get that $\D{\chi}/\chi\sim\left(n\sigma^{3}\right)\sim\phi$,
where $n=\rho/m$ is the number density and $\phi$ is the packing
fraction of the particles. We see from this estimate that the enhancement
due to fluctuations is stronger for dense gases and is strongest for
liquids.

However, the logarithmic divergence in (\ref{eq:chi_eff_2D}) means
that the contribution due to hydrodynamic fluctuations dominates for
sufficiently large (quasi) two-dimensional systems, regardless of
the density,\begin{equation}
\frac{\D{\chi}}{\chi}\sim\frac{k_{B}T}{\rho\chi(\chi+\nu)L_{z}}\,\ln\frac{L_{x}}{L_{\text{mol}}}\sim\left(n\sigma^{3}\right)\frac{\sigma}{L_{z}}\,\ln\frac{L_{x}}{\sigma}.\label{eq:dchi_est_2D}\end{equation}
A glance at Fig. \ref{fig:j1_vs_Nx} shows that the enhancement we
measured is only a few percent of the kinetic theory value (for our
DSMC simulations, $n\sigma^{3}=0.06$), and reaching the system width
$L_{x}$ where $\D{\chi}\sim\chi$ is impractical with DSMC simulations
at the present. In fact, for the parameters used in typical DSMC applications
the enhancement of the transport coefficients relative to the Chapman-Enskog
values is very small. Specifically, assuming an ideal hard-sphere
gas collision model and taking $L_{\text{mol}}=\D x_{c}$, for a quasi
two-dimensional system of thickness $L_{z}=\lambda$ we obtain the
estimate\[
\frac{\D{\chi}}{\chi}=\frac{16}{33\pi^{2}\left(n\lambda^{3}\right)}\,\ln N_{x}\approx\left(20N_{\lambda}\right)^{-1}\,\ln N_{x},\]
where $N_{\lambda}=n\lambda^{3}$ is the number of particles per cubic
mean free path, and $N_{x}=L_{x}/\D x_{c}$ is the number of collision
cells along the direction perpendicular to the gradient. In a typical
DSMC simulation $N_{\lambda}\approx100$, giving $\D{\chi}/\chi\approx0.5\cdot10^{-3}\,\ln N_{x}$,
which is less than $0.5\%$ even for $N_{x}=100$. While using molecular
dynamics instead of DSMC allows one to reach larger densities and
thus enlarge $\D{\chi}/\chi$, the regime in which $\D{\chi}\gtrsim\chi$
is difficult to access even using hard-disk MD \citet{VACF_2D_HS}
(see Fig. \ref{fig:D_HS2D}).

In this work we focused on the correlations between velocity and concentration
fluctuations. Concentration fluctuations also have long ranged self-correlations
$\sim k^{-4}$ in the presence of a concentration gradient, see Eq.
(\ref{eq:S_c_c}). Even though the enhancement of the concentration
fluctuations is proportional to the \emph{square} of the concentration
gradient, a two-dimensional calculation \citet{GiantFluctuations_ThinFilms}
similar to one presented here {[}see Eqs. (\ref{eq:dcdv_realspace},\ref{eq:j_lnLx}){]}
leads to the remarkable result,\[
\av{\delta c\left(\V r,t\right)\d c\left(\V r,t\right)}_{\text{neq}}^{(2D)}=\frac{3k_{B}T}{128\pi^{3}\rho\chi(\chi+\nu)L_{z}}\,\left(\frac{L_{x}}{L_{y}}\right)^{2}\,\left(\D c\right)^{2},\]
where $\D c=(\nabla\bar{c})L_{y}$ is the macroscopic concentration
variation. Assuming $L_{x}\sim L_{y}$, we thus obtain {[}c.f. Eq.
(\ref{eq:dchi_est_2D}){]}\[
\frac{\av{(\delta c)(\d c)}_{\text{neq}}^{(2D)}}{\left(\D c\right)^{2}}\sim\frac{k_{B}T}{\rho\chi(\chi+\nu)L_{z}}\sim\left(n\sigma^{3}\right)\frac{\sigma}{L_{z}},\]
where $n\sigma^{3}\sim1$ for liquids. We thus see that for systems
a few molecules thick, the non-equilibrium concentration fluctuations
can become comparable to the deterministic variation. In this case
we expect that a perturbative approach based on the linearized theory
will not be applicable and the use of a nonlinear finite-volume solver
will be indispensable.

\end{appendix}


\end{document}